\documentclass[12pt]{article}
\usepackage[numbers]{natbib}
\usepackage{graphicx,wrapfig,lipsum,amssymb,amsfonts,latexsym,amsmath,amsthm,times}
\usepackage[numbers]{natbib}
\usepackage{subfigure,epsfig}
\usepackage{morefloats}
\usepackage{amsbsy}
\usepackage{soul}

\usepackage{amsmath}
\usepackage{amssymb}
\usepackage{amsthm}
\usepackage{graphicx}
\usepackage{caption}
\usepackage{lipsum}
\usepackage{float}
\usepackage{upgreek}
\usepackage{array,multirow}
\usepackage{framed}
\usepackage{afterpage}
\usepackage{lscape}
\usepackage{color}
\restylefloat{table}
\usepackage{footmisc}
\usepackage{textcomp}
\usepackage{amsmath}
\usepackage{amssymb}
\usepackage{amsthm}
\usepackage{authblk}
\usepackage{bm} % Support for extra bold math symbols
\usepackage{booktabs} % Nicer tables
\usepackage{microtype} % Better font rendering, makes everything
\usepackage{float} % [H] alignment option for floats
\usepackage{textcomp} % Copyright symbol

\usepackage{algorithm,algpseudocode}
\DeclareMathOperator*{\argmin}{arg\,min}

\usepackage{float}

\floatstyle{plaintop}
\restylefloat{table}

\fontencoding{OT1}
\fontfamily{ptm}  % cmr/cmss/cmtt & Computer Modern Roman/Sans/Typewriter
\fontseries{m}    % m=Medium, b=Bold, bx=Bold extended, sb=Semi-bold, c=Condensed
\fontshape{n}     % n=Normal, it=Italic, sl=oblique, sc=small caps
\fontsize{11}{11}
\selectfont

\newcommand{\Up}{\Upsilon}
\newcommand{\told}{{\pmb{\theta}}^{\textrm{old}}}
\newcommand{\pt}{{\pmb{\theta}}}
\newcommand{\ut}{{\pmb{\upsilon}}}
\newcommand{\tnew}{{\pmb{\theta}}^{\textrm{new}}}
\newcommand{\toldi}{{\theta}^{\textrm{old}}_i}

\newcommand{\pderiv}[1]{\ensuremath{\frac{\partial}{\partial #1}}}

\newcommand{\mylist}{\begin{list}{$\bullet$}{ \setlength{\topsep}{0cm}
\setlength{\itemsep}{0cm} \setlength{\parsep}{0cm}}}
\newcommand{\mylistend}{\end{list}}

\newcounter{task}
\newcounter{subtask}
\newcounter{subsubtask}

\begin{document}


%\title{Parameter inversion techniques for radiation detection problems}
\title{Hybrid optimization and Bayesian inference techniques for a non-smooth radiation detection problem}
\author[1]{R\u{a}zvan \c{S}tef\u{a}nescu \thanks{rstefan@ncsu.edu}}
\author[1]{Kathleen Schmidt}
\author[2]{Jason Hite}
\author[1]{Ralph Smith}
\author[2]{John Mattingly}

\affil[1]{Department of Mathematics, North Carolina State University, Raleigh, NC, USA}\affil[2]{Department of Nuclear Engineering, North Carolina State University, Raleigh, NC, USA}

\date{}
\maketitle

\begin{abstract}

 In this investigation, we propose several algorithms to recover the location and intensity of a radiation source located in a simulated $250 \textrm{ m } \times 180 \textrm{ m }$ block in an urban center based on synthetic measurements. Radioactive decay and detection are Poisson random processes, so we employ likelihood functions based on this distribution. Due to the domain geometry and the proposed response model, the negative logarithm of the likelihood is only piecewise continuous differentiable, and it has multiple local minima. To address these difficulties, we investigate three hybrid algorithms comprised of mixed optimization techniques. For global optimization, we consider Simulated Annealing (SA), Particle Swarm (PS) and Genetic Algorithm (GA), which rely solely on objective function evaluations; i.e., they do not evaluate the gradient in the objective function. By employing early stopping criteria for the global optimization methods, a pseudo-optimum point is obtained. This is subsequently utilized as the initial value by the deterministic Implicit Filtering method (IF), which is able to find local extrema in non-smooth functions, to finish the search in a narrow domain. These new hybrid techniques combining global optimization and Implicit Filtering address difficulties associated with the non-smooth response, and their performances are shown to significantly decrease the computational time over the global optimization methods alone. To quantify uncertainties associated with the source location and intensity, we employ the Delayed Rejection Adaptive Metropolis (DRAM) and DiffeRential Evolution Adaptive Metropolis (DREAM) algorithms. Marginal densities of the source properties are obtained, and the means of the chains' compare accurately with the estimates produced by the hybrid algorithms.

\end{abstract}

\begin{keyword}
Inverse problems; Simulated Annealing; Particle Swarm; Genetic Algorithm; Implicit Filtering;  Differential Evolution Adaptive Metropolis; Delayed Rejection Adaptive Metropolis;
\end{keyword}

%opening

%\author{}
%\maketitle

\section{Introduction}

Detecting properties of radiation material in an urban environment is a matter of national safety and security, and it requires fast solutions for accurate identification of such threats to the nearby population. To address this problem, one can apply a network of sensors capable of monitoring ionizing photons and registering the energy of impacting gamma rays. Additionally, an accurate model of detector response is required to provide information a sensor would read for a given radiation source inside the limited domain.

%Moreover, since measurements are known to be altered by instruments errors, it is important to define appropriate statistical models to track potential sources of errors.

The input reconstruction problem has been investigated since the 1960s due to its importance for military applications and environmental monitoring. Depending on the field of study, the literature refers to it as a parameter estimation problem \cite{morelande2007detection,morelande2009radiation,rao2008identification}, inverse problem \cite{isakov2006inverse}, or data assimilation problem \cite{shekhawat2013nuclear}. There are a number of difficulties inherent to the source identification problem. First, the source properties may not be uniquely determined by the observations, in which case are said to be unidentifiable. Second, the problem is usually ill-posed and often has to be regularized to obtain a reasonable approximation to the solution \cite{alpay2000model}.

Standard models for radiation transport include the Boltzmann equation \cite{shultisRadShi2000} and Monte Carlo models \cite{briesmeister2000report}. Both of these models require significant computational resources and are difficult to use in practice when solving source identification problems since numerous model evaluations are needed. Consequently, a number of researchers have designed alternative parametrizations and surrogate models. For example in \cite{king2010urban}, the authors modeled the threat as a point gamma source and employed a physics-based parametrization of gamma particle transport. A fast radiation transport model is also available as a component of Synth, a gamma-ray simulation code written by Pacific Northwest National Laboratory \cite{brennan2004radiation,hensley1996synth}. Another approach employed a Gaussian mixture \cite{morelande2009radiation} to model the radiation field.

Among the methods utilized to recover the information about the source properties, Bayesian techniques are popular. They include direct application of the Bayes' theorem \cite{chandy2010models,chandy2008networked,king2010urban} or Markov Chain Monte Carlo methods \cite{xun2011bayesian}. Other common methodologies employ maximum likelihood estimators \cite{chandy2008networked,chin2008accurate,king2010urban}, cumulative sum with triangulation \cite{king2010urban}, mean of estimates \cite{chin2008accurate} and Kalman filters \cite{shekhawat2013nuclear}.

Only a few studies explicitly targeted the source identification problem in an urban environment. A recent one focuses on a simulated $1.3\textrm{ km} \times 0.9 \textrm{ km}$ area of downtown Philadelphia \cite{king2010urban}. In this work, we propose three  methods relying on maximum likelihood estimators to find the properties of a possible radiation source in a simulated $250 \textrm{ m } \times 180 \textrm{ m }$ block in downtown Washington, DC. Based on information from the OpenStreetMaps database, we developed a 2D-representation where the buildings are represented as a set of disjoint polygons. We employed Boltzmann transport theory \cite{shultisRadShi2000} to construct a piecewise continuous differentiable parameterized response model by modeling the threat as a point source and taking into account only photons which travel directly from source to detector. By representing the radiation source with only three components, 2D location coordinates and intensity, we ensure that our problem is well-posed and that no regularization is needed. We refer to the 2D location coordinates and intensity as properties or components of the source. Also, by assuming that a collection of sensors is dispersed throughout the domain with its convex hull covering the majority of the domain, we ensure that the sought location and intensity of the source are identifiable.

Due to the non-smooth characteristics and multiple local minima of the negative log of the likelihood function, regular gradient-based optimization techniques are not directly applicable. Consequently, we propose three hybrid techniques, each combining a global optimization method and Implicit Filtering \cite{kelley2011implicit}, a local optimization technique appropriate for non-smooth objective functions. In the first step, we employ either Simulated Annealing \cite{kirkpatrick1983optimization,pincus1970letter}, Particle Swarm \cite{Kennedy_PS_1995} or Genetic Algorithm \cite{fogel1975adaptation} to identify a sub-domain where the source components lie. In the second stage, partially relying on finite difference gradients and Hessian information, we employ the Implicit Filtering to finish the search. Based on a heuristic strategy, we make use of early stopping criteria for the global optimization methods to avoid their costly final stages. Not surprisingly, there have been many previous efforts to decrease the cost of the global methods by combining them with local optimization techniques such as those reported in \cite{dos2007combining,dos2009novel,navon1990combined}.

By coupling a probabilistic acceptance rule with a simplified Genetic Algorithm \cite{storn1997differential}, \citet{braak2006markov} constructed a population Markov-Chain with a unique stationary distribution; this is similar to the proven convergence of the Delayed Rejection Adaptive Metropolis algorithm \cite{Haario}. We verify the proposed hybrid techniques by comparing their results to those from the Delayed Rejection Adaptive Metropolis (DRAM) and the DiffeRential Evolution Adaptive Metropolis (DREAM) \cite{Vrugt} algorithms, which provide estimates of the source location and intensity based on constructed posterior distributions. These Metropolis-Hastings-based algorithms are advantageous for dealing with potentially correlated parameters and because of their rigourous convergence properties.

The remainder of the paper is organized as follows. In Section \ref{sec:model:numerical}, we introduce the  response model relying on simplified Boltzmann transport theory, the domain geometry, and the statistical model employed in this investigation. The problem is formulated in Section \ref{sec:inverse_problem}. We describe the proposed hybrid techniques for solving the source localization problem in Section \ref{sec:hybrid_techniques} and the Bayesian algorithms in Section \ref{sec:Bayesian_techniques}. In Section \ref{sec:numerical_experiments}, we present several numerical experiments describing their performance based on computational time (CPU) and accuracy. We draw conclusions in Section \ref{sec:Conclusion}. The algorithms associated with the inference methods used in this manuscript are summarized in the Appendix.

\section{Radiation transport model} \label{sec:model:numerical}

A complete description of gamma transport phenomena, derived from
Boltzmann transport theory, is given by
\begin{equation}
    \begin{split}
        \frac{1}{c}\pderiv{t}I(\mathbf{r}, E,\hat{\bm{\Omega}}, t) + \hat{\bm{\Omega}} \cdot \nabla I(\mathbf{r}, E, \hat{\bm\Omega}, t) + \Sigma_T(\mathbf{r}, E, \hat{\bm\Omega}, t) I(\mathbf{r}, E, \hat{\bm\Omega}, t)  ~~~~~~~~~~~~~~~~~~~~~~~~~~~~~~~~~~~~~~~\\
       = S(\mathbf{r}, E, \hat{\bm\Omega}, t) + \int_0^\infty dE' \int_{4\pi} \; d\hat{\bm\Omega}' \frac{E}{E'} \Sigma_s(\mathbf{r}, E'\to E, \hat{\bm\Omega}' \to \hat{\bm\Omega}, t) I(\mathbf{r}, E', \hat{\bm{\Omega}}', t).
    \end{split}
    \label{eq:boltzmann}
\end{equation}
Here $I$ and $S$ denote the gamma intensity per unit area and external gamma source in the
medium characterized by the position vector $r$, energy $E$, unit vector in the
direction of the gamma $\hat{\bm\Omega}$, and time $t$. The parameters
include the total macroscopic cross-section for gamma interactions $\Sigma_T$, the
double-differential macroscopic scattering cross-section $\Sigma_s$, and the
speed of light $c$. We refer readers to \citet{shultisRadShi2000} for a more
detailed treatment of transport theory.

The problem of inferring the radiation source location and intensity from sensor measurements requires
the evaluation of the radiation transport model \eqref{eq:boltzmann} at various points in the
feasible space. The solution of the Boltzmann equation \eqref{eq:boltzmann} is typically quite
computationally demanding. It is often not feasible for inverse and uncertainty quantification methods requiring numerous model simulations.

Instead, we will employ a model that only considers gamma rays that travel
directly from source to detectors, without taking into account photons that suffer collisions. This approach relies on the assumption that photons
undergoing interactions in the medium have a very small probability of ever
arriving at a detector. We also assume that the physical scale of our problem is sufficiently large
so that both the source and detectors can be localized to points inside the domain. We will denote the location of the source as
$\mathbf{r}_s$ and associated intensity by $S_0$, which we assume is low enough
that we can ignore time-dependent effects. Under these assumptions, equation \eqref{eq:boltzmann} can be simplified to
\begin{equation}
    \hat{\bm{\Omega}} \cdot \nabla I(\mathbf{r}, E, \bm\Omega) + \Sigma_T(\mathbf{r}, E, \hat{\bm{\Omega}}) I(\mathbf{r}, E, \hat{\bm\Omega}) =
    \frac{S_0}{4\pi} \delta(E - E_0) \delta(\| \mathbf{r} - \mathbf{r}_s \|_2).
    \label{eq:noscatter}
\end{equation}
See \cite{shultisRadShi2000} for more details. Equation \eqref{eq:noscatter} can be solved to determine the intensity of photons arriving at any point $\mathbf{r}$ inside domain. This enables the computation
of the count rate measured by the $i$-th detector $D_i$
assuming that detectors are point detectors with face area $A_i$ and dwell
time $\Delta t_i$. The detector intrinsic efficiency $\epsilon_i\in [0,1]$ is
usually known in practice. If the $i^{\textrm{th}}$ detector is located at
point $\mathbf{r}_d^i$, the solution
\begin{equation}
    \hat{f}(D_i,\pt) = S_0 \Delta t_i \cdot \epsilon_i \cdot \frac{A_i}{4\pi \| \mathbf{r}_d^i - \mathbf{r}_s \|^2_2}
        \cdot \exp\left( \int_{\mathbf{r}_d^i - \mathbf{r}_s} \Sigma_T \; d\mathbf{s} \right)
    \label{eq:analytic}
\end{equation}
of equation \eqref{eq:noscatter} predicts the number of counts observed by the sensor given the location and intensity of
the source $\pt=(\mathbf{r}_s,S_0)$. The derivation of model response \eqref{eq:analytic} follows in a manner similar to that shown in \citet[Chapter 10.1.3]{shultisRadShi2000}, where the resulting solution is evaluated at the detector location $\mathbf{r}_d^i$.

\subsection{Model geometry, detectors' locations, and physical properties }\label{sec:model:physical}

\begin{figure}[t!]
 \centering
    \includegraphics[width=10.5cm]{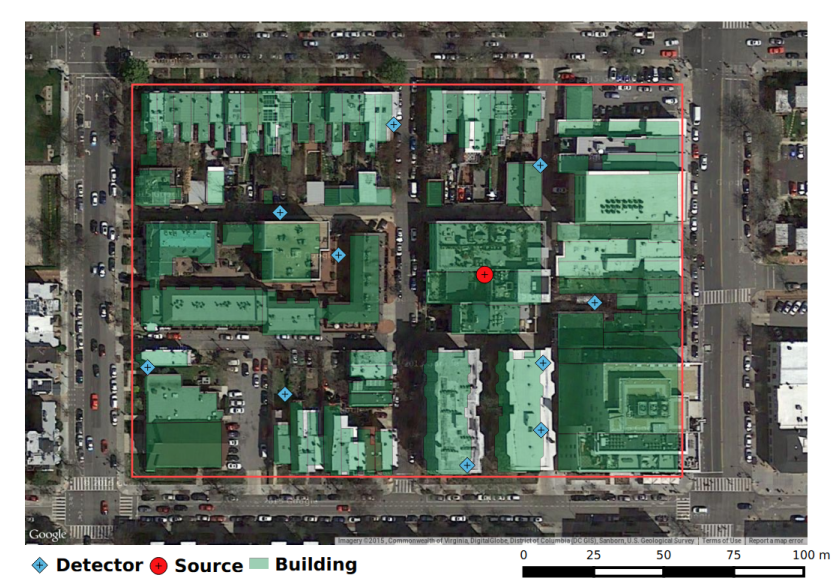}
    \caption{Satellite image of problem domain with model geometry overlaid\protect\footnotemark.}
    \label{fig:geometrymap}
\end{figure}

To provide a realistic example of an urban area, we randomly selected a $250 \textrm{ m } \times 180 \textrm{ m }$ block in downtown Washington, D.C., located at approximately
$38^{\circ}\,54'\,48''\,N$ by $77^{\circ}\,1'\,60''\,W$ (Johnson Avenue NW) to serve as our domain.
Buildings in this area are primarily brick and concrete
residential housing and are generally $1-5$ stories in height. Using data from
the OpenStreetMaps database (\texttt{https://www.openstreetmap.org/}), we constructed a 2D representation of the area
to serve as the test geometry. Our implementation treats the buildings as a set of disjoint polygons $P_j,~j=1,2,..,N_g$, each of which is assigned a corresponding macroscopic cross-section. This is equivalent to simultaneously
specifying the composition and density of the material for a fixed volume. A satellite photo of the area with an overlay of the constructed representation is provided in Figure \ref{fig:geometrymap}.
\footnotetext{Imagery \textcopyright 2016 Commonwealth of Virginia, DigitalGlobe,
District of Columbia (DC GIS), Sanborn, U.S. Geological Survey, Map data
\textcopyright 2016 Google}

%\begin{wrapfigure}{r}{7.5cm}
%    \includegraphics[width=7.5cm]{Figures/revised-map.png}
%    \caption{Satellite image of problem domain with model geometry overlaid\protect\footnotemark}
%    \label{fig:geometrymap}
%\end{wrapfigure}
%\footnotetext{Imagery \textcopyright 2016 Commonwealth of Virginia, DigitalGlobe,
%District of Columbia (DC GIS), Sanborn, U.S. Geological Survey, Map data
%\textcopyright 2016 Google}
%

Approximate calculations indicate that wood and concrete buildings would correspond to an
optical thickness of around $3$ mean free paths (MFPs). The mean free path denotes the mean distance travelled by the ionizing photons between collisions with atoms of the building. Consequently, we randomly selected cross-sections for each building so that their optical thickness would be between $1$
and $5$ MFPs.  The random sampling was also weighted according to the
volume of each building, so that smaller buildings were biased towards
smaller optical thicknesses and vice versa. The regions between buildings were treated as dry
air at standard temperature and pressure, with cross-sections taken from the
NIST XCOM database (\texttt{http://www.nist.gov/pml/data/xcom/}).

Sampling from a uniform distribution, we generated the locations of $10$ detectors in the domain denoted by diamond marks in Figure \ref{fig:geometrymap}. The specific dispersal pattern was selected to spread the detectors evenly
throughout the area. We assumed that detectors had areas $A_i$ of $3$-inches diameter by $3$-inches length for incident gamma energy of $662\;\text{KeV}$ which is standard packaging for  NaI scintillators, with detectors absolute efficiency of $\epsilon_i = 62\%,~i=1,2,..,10$. The dwell time $\Delta t_i,~i=1,2,..,10$, for all detectors was chosen to be $1\;\text{seconds}$ (s).  For the background, we took the nominal intensity to be $B = 300\text{~counts per second}$ (cps) for the entire domain, which is typical for a $3'\times 3'$ NaI detector in the U.S. Southeast.

\subsection{Numerical model for detector response} \label{sec:model:implementation}

To determine the intensity of photons arriving at a given detector location
$\mathbf{r}_d^i$, the algorithm employs a simple ray-tracing scheme. Starting
at the location of the source, we draw a ray from $\mathbf{r}_s$ to
$\mathbf{r}_d^i$. We then compute the intersection of this ray with the disjoint polygons $P_j,~j=1,2,..,N_g,$ representing the set of buildings in our domain. This yields a series of line segments expressing the
path traversed in each region. If we assume a given ray intersects $N_\ell$ polygons, $N_\ell<N_g$,
and let $\mathcal{L}=\{ (\ell_j, \Sigma_T^{(j)})
\}_{j=1}^{N_\ell}$ be the set of all intersecting segments, where $\mathbf{\ell}_j$
is the Euclidean length of the $j$-th segment and $\Sigma_T^{(j)}$ is the
corresponding value for the macroscopic total cross-section,
then equation \eqref{eq:analytic} takes the form
\begin{equation}
    \hat{f}(D_i,\pt) = S_0 \Delta t_i \cdot \epsilon_i \cdot \frac{A_i}{4\pi \| \mathbf{r}_d^i - \mathbf{r}_s \|^2_2}
    \exp\left( -\sum_{j=1}^{N_\ell} \ell_j \cdot \Sigma_T^{(j)} \right).
    \label{eq:discrete}
\end{equation}
%\footnote{\texttt{https://pypi.python.org/pypi/Shapely}}

Equation \eqref{eq:discrete} provides an analytic expression approximating the
detector response, and its primary computational pertains to calculating the
intersection of lines with the model geometry. Equation \eqref{eq:discrete} represents a dramatic
simplification to the solution of \eqref{eq:boltzmann}, a non-linear PDE with
seven independent variables whose solution in complex geometries can require
many hours even on a supercomputer. We implemented the numerical model
\eqref{eq:discrete} as a short Python code. It employs the Shapely library
({\texttt{https://pypi.python.org/pypi/Shapely}}) for performing the computational
geometry calculations. The model takes as input a specification of polygons
representing the different regions of the domain, cross-section data, detector
locations, source intensity and source location.

\subsection{Statistical Model}

To generate syntectic data, we chose a radiation source with an activity of $87$ millicuries$\text{ (mCi) } = 3.219$ gigabecquerels (GBq), which is equivalent to
$1\;\text{mg}$ of cesium-137 (Cs-137). Ray tracing calculations were performed in terms of photon mean free paths (MFPs), with our point of reference being a $\gamma$
energy of $662\;\text{keV}$, resulting from the $\beta^-$ decay of Cs-137.
The source location was picked at random, albeit constrained to lie near the
center of the domain, with the coordinates $(158, 98) \;\text{m}$. Consequently, the location and intensity of the source is
\begin{equation*}
  \pt_0 = ({\mathbf{r}}^0_s,S_0^0) = \left((158, 98),87 \text{mCi}\right).
\end{equation*}
%

%and model errors $\delta$
To construct a statistical model, we consider a uniform background with nominal intensity $B$. It is well known that radioactive decay and detection are Poisson random processes. By including Poisson random effects, we produce an initial statistical model
\begin{equation}\label{Victoras}
  \Up_i \sim \mathtt{P}\left( \hat{f}(D_i,\pt_0) + B\right)
\end{equation}
associated with the $i^{\textrm{th}}$ detector response. The Poisson distribution of mean $f(D_i,q) = \hat{f}(D_i,\pt_0) + B$ is denoted by $\mathtt{P}$. In this way, we naturally model the observations associated with each detector as random variables $\Up_i,~i=1,2,..,10$.

\section{Inverse problem description} \label{sec:inverse_problem}

The problem of detecting the location and intensity of a radiation source when several detectors and associated measurements are available represents a classical inverse problem.  To address this, we will explore various methods as well as the central role of the likelihood function.

As discussed in previous section, we generated synthetic responses at $10$ detector locations using a given location and intensity of the radiation source $\pt_0$. To test the accuracy of the inverse algorithms, we assume that $\pt_0$ is unknown and we infer it from realizations $\ut_i \in \mathbf{R}^{10} $ of the random variables $\Up_i,~i=1,2,..,10$.

The Poisson likelihood function $L : \Omega \to [0,\infty)$ is given by

\begin{equation}
\label{eqn:Poisson_Likelihood}
\pi(\pmb{V}|\pt)= \prod_{i=1}^{10} \frac{f(D_i,\pt)^{\sum_{j=1}^{10}v_{i,j}}\cdot e^{-10 f(D_i,\pt)}}{\upsilon_{i,1}!\upsilon_{i,2}!\cdot \cdot \cdot \upsilon_{i,10}!},
\end{equation}
where $\Omega = [0,250] \times [0,180] \times [5\cdot 10^8,5 \cdot 10^{10}]$ and $\pmb{V} = [\ut_1,\ut_2,..,\ut_{10}]$ is the vector of all the available observations.

We employ two strategies in this paper.  Initially, we will focus on developing and implementing efficient algorithms capable of accurately estimating the location and intensity of the source. We subsequently employ Bayesian techniques to construct distributions for the source properties.

\section{Stochastic and deterministic hybrid techniques} \label{sec:hybrid_techniques}

\begin{figure}[b!]
  \centering
  \subfigure[] {\includegraphics[scale=0.35]{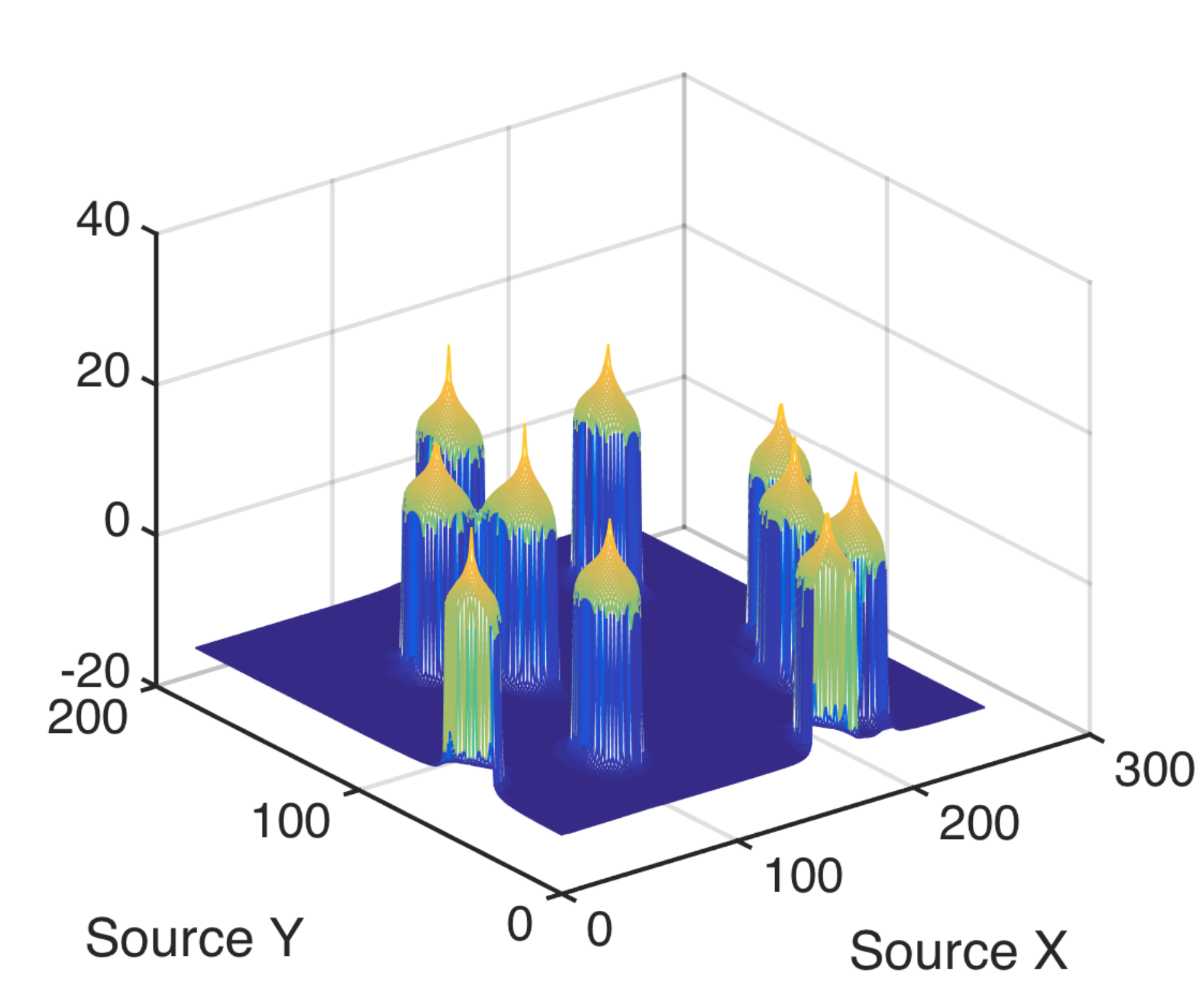}}
  \subfigure[]{\includegraphics[scale=0.35]{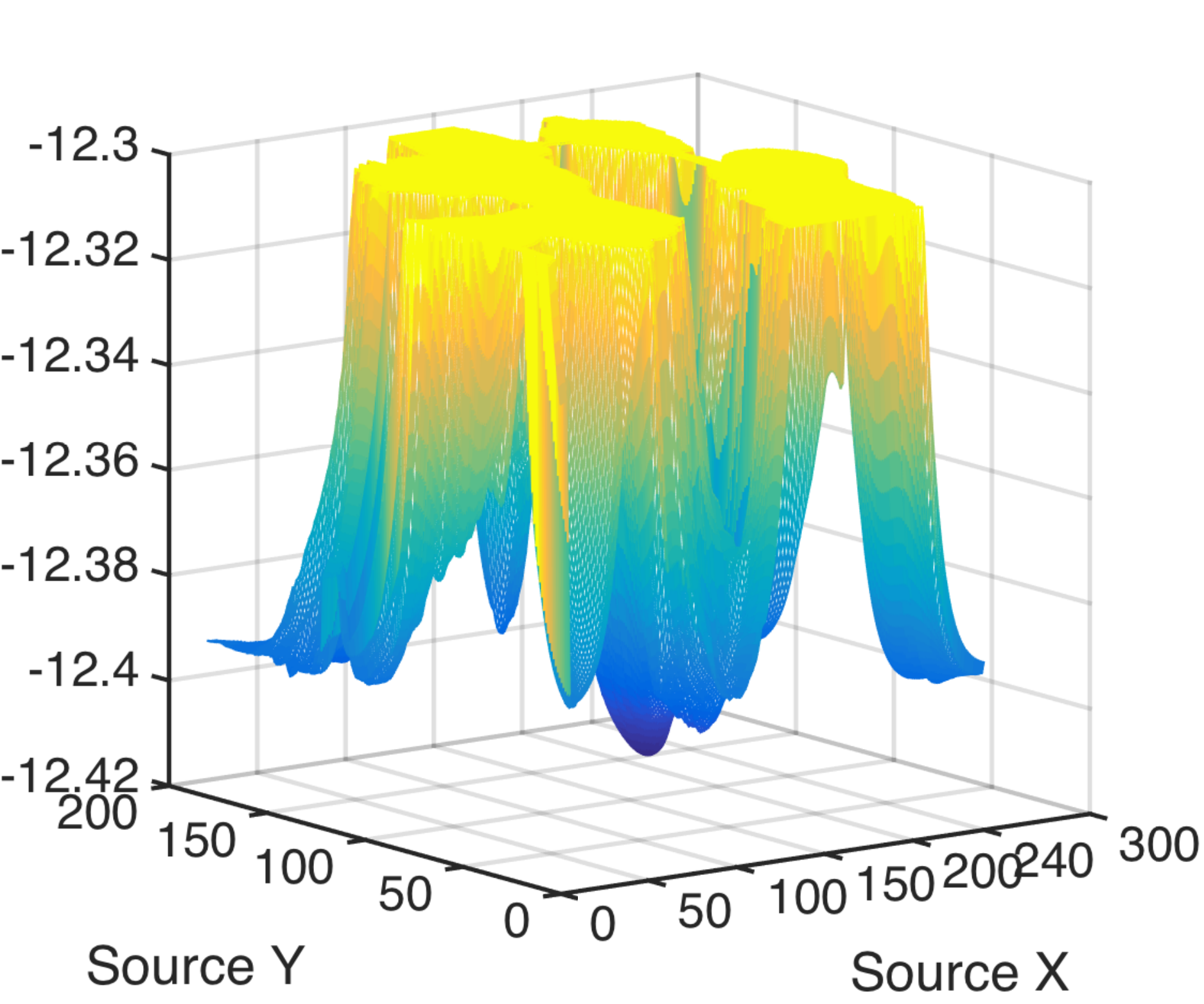}}
\caption{Negative logarithm of the Poisson likelihood: (a) full, (b) cross section.}
\label{Fig::Log_Poisson_Likelihood}
\end{figure}

One technique to estimate the location and intensity of a radiation source based on measured data is to apply maximum likelihood estimators. Due to the monotonicity of the logarithm function, maximizing \eqref{eqn:Poisson_Likelihood} is equivalent to minimizing the negative logarithm of the likelihood and the maximum likelihood estimate is obtained by solution
\begin{equation}
\label{eqr:neg_log_poisson_likelihood}
 \min_{\pt \in \Omega}  J(\pt),~ J(\pt)  =  \frac{1}{2} \sum_{i=1}^{10}\sum_{j=1}^{10}\big[-v_{i,j}\cdot \log(f(D_i,\pt))+f(D_i,\pt)\big].
\end{equation}
However, due to the complex nature of the domain geometry, classical optimization techniques can not be applied since the objective function has multiple local minima and discontinuities. Figure \ref{Fig::Log_Poisson_Likelihood} depicts the negative log-likelihood Poisson function \eqref{eqr:neg_log_poisson_likelihood} for a radiation source of $1\;\text{mg}$ of Cs-137; i.e., a nominal intensity of $S_0^0 = 3.2\times 10^9$ decays per second. Note that the $10$ peaks observed in Figure \ref{Fig::Log_Poisson_Likelihood}(a) are correlated with the detectors' locations. Moreover the function's floor is not smooth as illustrated in the zoomed perspective in Figure \ref{Fig::Log_Poisson_Likelihood}(b).

\begin{figure}[t!]
\centering
%\begin{wrapfigure}{r}{7.5cm}
\includegraphics[trim=0.2cm 8cm 0.5cm 0cm,width=10.5cm]{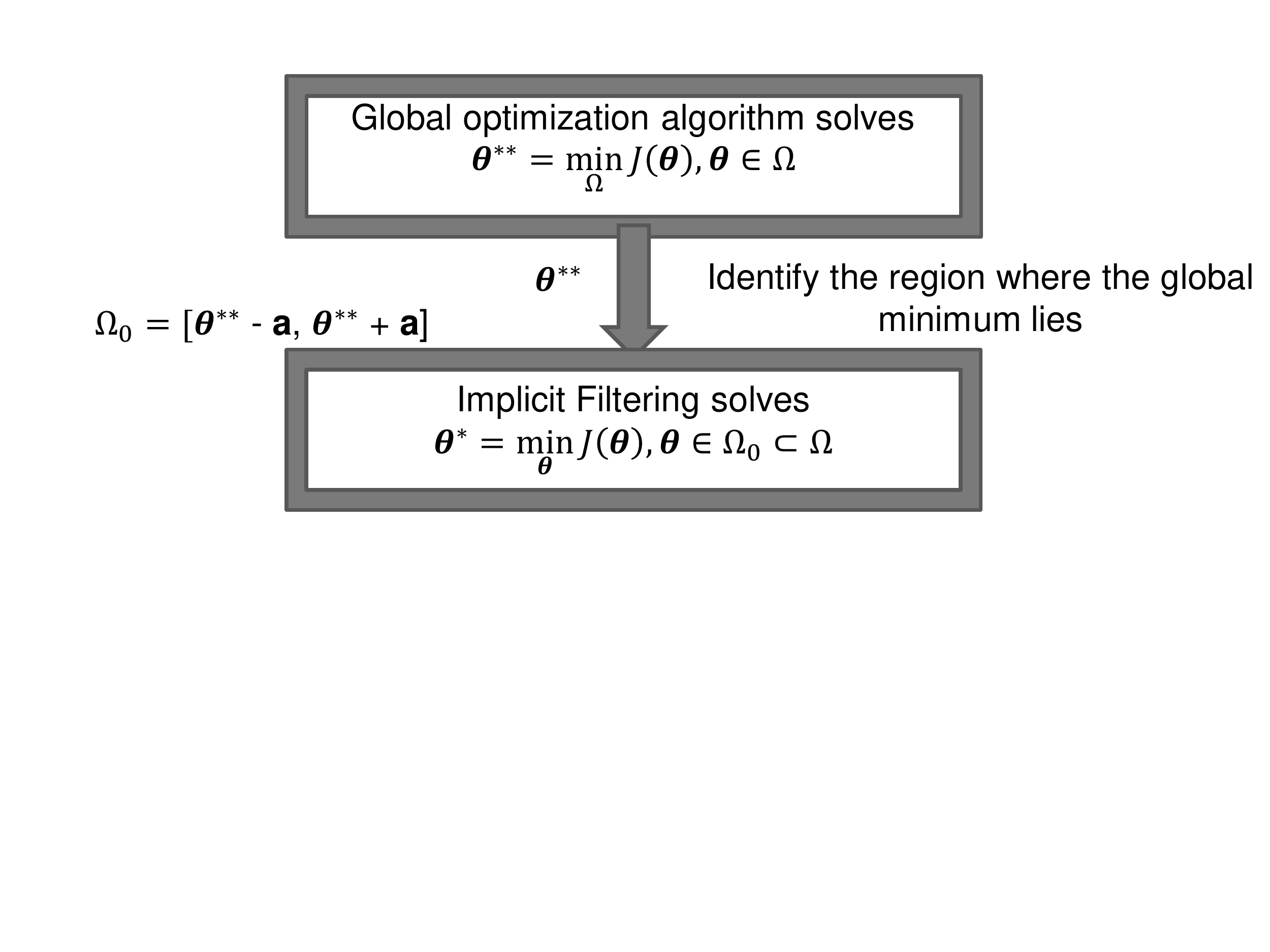}
\caption{Mixed optimization algorithm for the non-smooth radiation inverse problem.}
\label{Fig::Mixed_Optimization}
\end{figure}

Due to the non-smooth and non-convex nature of $J(\pt)$, we propose a hybrid strategy based on a combination of global and local optimization techniques. In the first stage, we will employ a global optimization method to identify a region of the domain where the global minimum lies. This is followed by a second stage where we use a local  optimization method specialized for non-smooth cost functions to rapidly identify the source properties inside a narrow domain. The concept is summarized in Figure \ref{Fig::Mixed_Optimization}. For the global optimization techniques, we will employ stochastic and heuristics approaches, which rely solely on objective function evaluations. As for the local technique, we employ Implicit Filtering \cite{kelley2011implicit}, which utilizes a coordinate search and finite difference approximations of the gradient and Hessian to finish the search. The usage of the terms gradient and Hessian is in a loose sense since they may not even exist.

%\end{wrapfigure}
%

These hybrid algorithms exploit the ability of global methods to find global minima points even when the objective function is non-smooth. At the same time it makes use of the Implicit Filtering technique, which is able to quickly and accurately determine local optima points. By utilizing early stopping criteria for the global methods, we avoid refining the search in the proximity of the optimum known for its extensive computational complexity. Challenges of the proposed mixed optimization strategy include identification of stopping criteria for global optimization techniques, and selection of parameter $\pmb{a}$ and narrow sub-domain $\Omega_0$ described in Figure \ref{Fig::Mixed_Optimization}.

\subsection{Global optimization algorithms}

Non-convex problems, such as that defined in \eqref{eqr:neg_log_poisson_likelihood}, have many different local optima points, thus making it difficult to determine a global minimum. Furthermore, the objective function is discontinuous with respect to the source location $\mathbf{r}_s$ due to the response radiation model and the domain geometry.

In the recent years, a number of global methods based solely on function evaluation have been developed. The majority are based on natural phenomena analogies, and try to mimic the efficiency and simplicity of natural optimized processes. Algorithms motivated by species evolution \cite{fogel1975adaptation,goldberg1989genetic}, annealing of metals \cite{kirkpatrick1983optimization,pincus1970letter}, neurons' inter-communication
\cite{cochocki1993neural}, ants' social behavior \cite{bonabeau2000inspiration},
and immune cells behaviour \cite{de2001ainet} have been proposed and applied to solve difficult optimization problems. Among them, perhaps the most efficient techniques are
Simulated Annealing \cite{pincus1970letter}, Particle Swarm \cite{Kennedy_PS_1995} and Genetic Algorithms  \cite{fogel1975adaptation} with the last two being part of a more general class of evolutionary algorithms. %\cite{storn1996minimizing}.

\subsubsection{Simulating Annealing (SA)}

Simulated Annealing is a stochastic technique for approximating the global optimum of a given function. It is based on a Metropolis algorithm \cite{metropolis1953equation}, initially used to compute efficient simulations of a collection of atoms in equilibrium at a given temperature. Later the Metropolis algorithm was modified to include a temperature schedule \cite{kirkpatrick1983optimization} to facilitate searching optimal wiring configurations in a densely wired computer chip with associated non-convex objective-functions. Simulated Annealing can be described  as a sequential search procedure. It can escape from local minima by accepting transitions associated to an increase in the cost function value in addition to transitions corresponding to a decrease in objective function value.

\begin{figure}[t!]
  \centering
\includegraphics[scale=0.31]{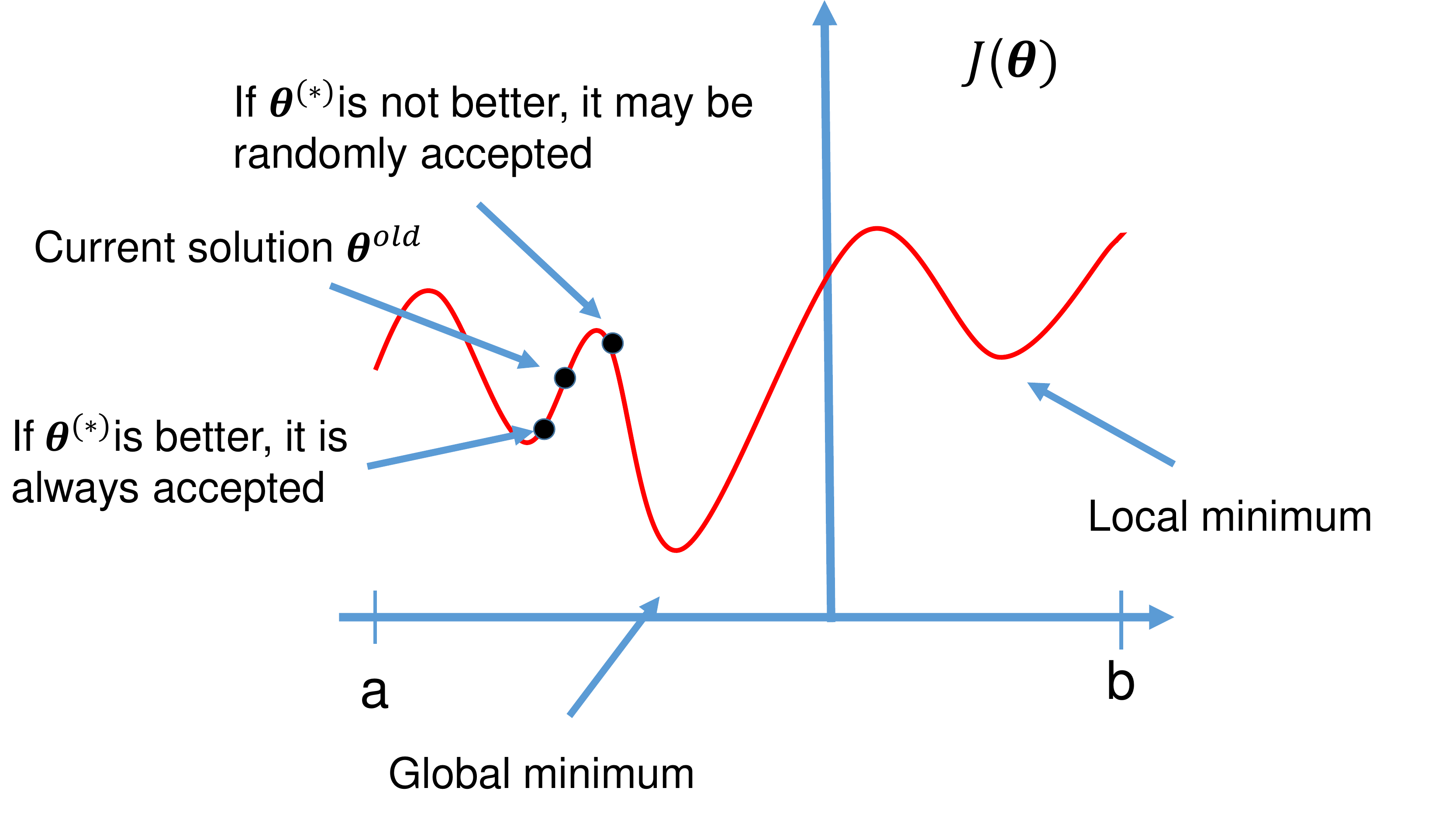}
\caption{Simulated Annealing Description}
\label{Fig::SA_description}
\end{figure}

An intuitive description of the method is presented in Figure \ref{Fig::SA_description}. The uphill samples are accepted in a limited way depending on a probabilistic acceptance criterion. In the course of the minimization process, the probability of accepting uphill points descends slowly towards zero based on a decreased temperature schedule choice. Accepting points that increase the objective function value helps to explore the state space entirely.

The Simulated Annealing algorithms can be classified into two main subcategories: continuous and discrete, depending on the nature of the search space of the optimization problem under consideration. The majority of the discrete Simulated Annealing studies focused on discrete combinatorial optimization problems \cite{aarts1989simulated,aarts1988simulated,kirkpatrick1983optimization}. In this discrete case, \citet{geman1984stochastic} and \citet{hajek1988cooling} have provided necessary and sufficient conditions for cooling temperature schedules that guarantee convergence in probability to the global optimum. Continuous Simulated Annealing algorithms have also been proposed by \citet{bohachevsky1986generalized,corana1987minimizing,dekkers1991global}. In contrast to discrete Simulated Annealing, \citet{romeijn1994simulated} showed that the sequence of sample points converges in probability to a global optimum regardless of how fast the temperatures converge to zero.

We note that convergence results have been obtained solely for continuous objective functions.
For piecewise continuous negative log-likelihood functions such as ours, convergence results are not available. The extension of the convergence results for Lebesgue integrable density functions is straightforward. Consequently we employ an adaptive Simulated Annealing algorithm \cite{chen1999adaptive,ingber1992genetic} available in MATLAB. An iteration of the adaptive Simulated Annealing is described in the Appendix \ref{sec:Appendix_SA}.

To increase the efficiency of the adaptive Simulated Annealing scheme, a simple multi-starting strategy \cite{malek1989serial} is proposed. The basic idea is to run several Simulated Annealing algorithms in parallel and periodically compare the results.  Initially, a predefined number $P$ of starting states points are uniformly selected from uniform distributions $U(0,250),~U(0,180),~U(5\cdot 10^8,5\cdot 10^{10})$ and $P$ separated Simulated Annealing threads are launched in parallel. As soon as one Simulated Annealing run satisfies one of the stopping criteria, the multi-starting algorithm stops. In our numerical experiments, this approach has been proved to be effective in avoiding the local minima in comparison with a serial version of Simulated Annealing based solely on a single thread. While being faster, the parallel version increases the required number of cost function evaluations.

\subsubsection{Particle Swarm (PS)}

The Particle Swarm approach is a global, metaheuristic optimization algorithm motivated by social-psychological principles \cite{kennedy2010particle}. It was originally introduced by \citet{Kennedy_PS_1995} and was designed to imitate a social behaviour such as the movements of birds in a flock or fishes in a shoal. Later the algorithm was simplified and its performance for solving optimization problems were reported in \cite{eberhart1995new}.

Whereas there is no available convergence theory, the Particle Swarm method has been shown to efficiently determine the solutions of a wide range of global optimization problems associated with smooth \cite{angeline1998evolutionary,trelea2003particle} and non-smooth \cite{eberhart1998evolving,park2005particle} objective functions. Since the method does not require a differentiable objective function, it represents a suitable method for our inverse problem.

The method utilizes a population of particles that generate interconnecting trajectories inside the search-space
according to stochastic mathematical formulae over the particles' position and velocity. Each particle's movement is influenced by its local best known position associated with the smallest objective function value. At the same time, it is also guided toward the best known positions revealed by other particles. Whereas the Particle Swarm terminology uses particles to identify state points, here we will employ the later nomenclature for a uniform description of the optimization techniques. The Particle Swarm is naturally parallelizable and it is available in MATLAB. A description of the algorithm is provided in Appendix \ref{sec:Appendix_PS}.

\subsubsection{Genetic Algorithm (GA)}

Introduced by \citet{fogel1975adaptation}, the Genetic Algorithm is another popular heuristic method used to generate  solutions to global optimization problems. It mimics the process of natural evolution and terms such as selection, elitism, crossover, and mutation are employed to denote proposal schemes for new state points along the search procedure. The Genetic Algorithm is similar to the Particle Swarm method in the sense that both evolutionary techniques are population-based search methods relying on multiple trajectories to explore the feasible space. However, their proposal functions differ significantly.

Here we follow a direct value encoding and a global parallelization implementation \cite{cantu1998survey} where the evolution of the new state points and the application of the genetic operators are explicitly parallelized. The method implementation is available in MATLAB and usually produces a speedup proportional to the number of processors utilized during the simulations. The algorithm is summarized in Appendix \ref{sec:Appendix_GA}.

In its present form, there is no theoretical guarantee the Genetic Algorithm will converge to the global optimum. When coupled with probabilistic acceptance rules, new Bayesian methods emerged \cite{braak2006markov,vrugt2009accelerating}. They have good theoretical convergence results providing information about the entire parameters distributions. This comes with additional computational costs and in cases where fast estimates are desired, the standard Genetic Algorithms should be employed.

\subsection{Local optimization strategies}

Local optimization techniques typically employ the gradient to explore the feasible space and search for optimal points. The associated functions must be continuously differentiable, and for convex functions the local techniques quickly converge to the unique minimum point. There have been several investigations to extend the classic theoretical framework to accommodate locally Lipschitz continuous objective functions. Non-smooth optimization techniques exploit the subdifferential theory developed by \citet{rockafellar2015convex} and \citet{clarke1990optimization}. Among them, the bundle techniques \cite{haarala2004new} are the most reliable. The idea is to approximate the subdifferential of the objective function by gathering subgradients from previous steps into the bundle. For our problem, due to the irregularity of the domain, it is difficult to compute the subgradients, so we need to rely on more general approaches such as sampling techniques.

\subsubsection{Implicit Filtering (IF)}

Implicit Filtering is a deterministic sampling method designed to solve local bound-constrained optimization problems. Since its introduction by \citet{winslow1991doping}, several versions \cite{choi1999iffco,gilmore1993iffco} and research studies \cite{choi2000superlinear,gilmore1995implicit} have been published. A comprehensive description of the method, together with asymptotic convergence results, are available in the book by \citet{kelley2011implicit}. The technique is capable of solving optimization problems for noisy, non-smooth discontinuous or random objective functions which may not even be defined at all points in the feasible space \cite{kelley2011implicit}.

The method seeks among the stencil points for lower objective function values following the coordinate search guidance. At the same time, it explores the feasible space using either a modified projected quasi-Newton or a Gauss-Newton methods depending on the type of the objective function. We describe only the quasi-Newton approach in Appendix \ref{sec:Appendix_IF} and refer readers to \cite{kelley2011implicit} for details regarding the Gauss-Newton method.

By incorporating Implicit Filtering inside our mixed optimization strategies, it allows us to locally refine the search in smaller sub-domains centered on the output states generated by the global techniques. Moreover, Implicit Filtering enjoys asymptotic convergence relying on the assumption that the objective function $J$ is a perturbation of a Lipschitz continuously differentiable function with sufficiently small noise. Our objective function satisfies these assumptions, and as a consequence of stencil failure, it was proved that the Implicit Filtering sequence converges to a local minimum \cite{kelley2011implicit} inside the feasible domain.
%\documentclass[11pt]{article}
%\topmargin-.5in
%\textwidth6.6in
%\textheight9in
%\oddsidemargin0in

%\section{Bayesian Parameter Estimation Techniques}
\section{Bayesian Inference Techniques} \label{sec:Bayesian_techniques}
To verify the hybrid techniques and to quantify the uncertainty of the source location and intensity, we employ two Bayesian techniques, the Delayed Rejection Adaptive Metropolis (DRAM) algorithm and the DiffeRential Evolution Adaptive Metropolis (DREAM) algorithm. These Bayesian methods provide marginal posterior distributions for each source property and hence allow us to quantify uncertainties associated with source properties.

\subsection{Delayed Rejection Adaptive Metropolis (DRAM)}

The DRAM method \cite{Haario,smith2014uncertainty} is a modified version of the Metropolis-Hastings algorithm, which includes adaptation and delayed rejection to improve mixing and exploration of the posterior. The adaptive step permits the geometry of the proposal function to be updated as new information about the posterior densities is acquired, and the delayed rejection step improves the mixing of the chains and provides improvements when the adaptation process has a slow start. Details are provided in Appendix \ref{sec:Appendix_DRAM}. The stochastic process resulting from the algorithm is non-Markovian. Based on an ergodicity result \cite{Haario}, the DRAM algorithm yields asymptotically unbiased estimators for expected values of the searched parameters or states with respect to their posterior distribution.

\subsection{DiffeRential Evolution Adaptive Metropolis (DREAM)}

Whereas the adaptation and delayed rejection components of DRAM are often sufficient to construct posterior parameter densities for a wide range of problems, there are some problems for which the algorithm is not efficient, especially those involving complex, multimodal, or heavy-tailed posteriors \cite{smith2014uncertainty}. For such applications, parallel chains were incorporated into adaptive Metropolis algorithms, resulting in differential evolution Markov chain methods.

The DiffeRential Evolution Adaptive Metropolis method is a generalization form of Differential Evolution Markov Chain \cite{braak2006markov} that utilizes a randomized subspace sampling strategy and a delayed rejection module. Moreover, the proposal function for the current chain state point may depend on more than $2$ additional chains iterates improving the efficiency of the search. More details are provided in Appendix \ref{sec:Appendix_DREAM}. Despite having multiple chains, DREAM can be viewed as defining a single stochastic process on the state space $\Omega^P$ \cite{Vrugt}, where $P$ is the number of chains. This stochastic process has good theoretical properties such as ergodicity and a unique stationary distribution.

  %\newpage

\section{Numerical Examples} \label{sec:numerical_experiments}

In Section \ref{sec::hybrid_exp}, we discuss the performance of the hybrid global-local algorithms. We also compare their results with the outputs of the Bayesian algorithms and discuss the relative efficiency and accuracy. In Section \ref{sec::Bayesian_exp}, we illustrate the manner in which the Bayesian algorithms can be used to construct posterior densities for the location and intensity of the radiation source.

As depicted in Figure \ref{fig:geometrymap}, we consider a simulated downtown area in Washington DC as our domain. The statistical model was described in equation (\ref{Victoras}), and for our simulations we generated $10$ measurements for each of the $10$ sensors uniformly distributed over the entire analytical domain $\Omega = [0,250] \times [0,180] \times [5\cdot 10^8, 5\cdot10^{10}]$.

\subsection{Hybrid Techniques} \label{sec::hybrid_exp}

We note that the proposed hybrid techniques are constructed by coupling global and local optimization techniques. The purpose is to reduce the computational burden of solely using global optimization techniques by taking advantage of the convergence properties of the local technique once the minimization search arrives in the proximity of the optimal point.

\paragraph{Example 1} We first run solely the global optimization techniques to motivate the need for hybrid methods. We select a population $P$ of $16$ points for all the methods initiated from uniformly randomly selected points inside the domain. In the case of SA method, this is equivalent to $16$ independent trajectories. Since the intensity of the source varies significantly, we scale it to span the interval [1,100]; i.e., $S_0 = S_0/(5 \cdot 10^8)$. Table \ref{tbl::global_param} includes the details regarding setting up the global optimization techniques. For SA, $r_p$ denotes the reannealing parameter, whereas $T_1^0,~T_2^0,~T_3^0$ are the source properties associated temperatures defining the initial sampling space \eqref{eq::SA_proposal_function}. For PS, $\textrm{Ns}$ represents the initial size of each particle neighborhood. The inertia parameter $W$ and the self and social adjustment variables $y_1$ and $y_2$ contribute to new velocity proposals \eqref{eqn::PS_velocity_formula}. For GA, $r_e,~r_c,~r_m$ define the number of elite, crossover and mutation fractions. The maximum number of function evaluations or model runs is fixed to $10,000$.
\begin{table}[t!]
\begin{center}
  \begin{tabular}{ |c | c | c| } \hline
    SA & PS & GA \\ \hline
    $T_1^0 = 240,~T_2^0 = 180,~T_3^0 = 99,~$ & $\textrm{Ns} = 4,~W = 1.1$ & $r_e = [P\cdot0.05]+1$   \\ \hline
    $r_p = 50$ & $y_1 = y_2 = 1.49$ & $r_c = 0.2*(P-r_e),~r_m = P-r_e-r_c.$ \\
    \hline
  \end{tabular}
  \end{center}
     \caption{Design parameters of global optimization methods for Simulated Annealing (SA), Particle Swarm (PS) and Genetic Algorithm (GA). See Appendices A-C for more details.}
     \label{tbl::global_param}
\end{table}

All of the techniques successfully identified the components of the radiation source and had similar errors. For the intensity, we computed the relative errors. We note in Table \ref{table::global_techniques_output_Exp1} that GA and PS exploit communicating trajectories and finished the search in only $1,232$ and $1,072$ model evaluations. This is not the case for the SA method, which requires $28,191$ model evaluations.  Since all of the employed methods are stochastic in nature, multiple simulations are required to verify the results. In $10,~9$, and $8$ cases out of $10$ simulations, respectively, the SA, PS and GA methods accurately estimated the properties of the source. This indicates that the SA method has better capability to avoid local minima than PS and GA. However, the performance of SA technique is achieved at a very high computational cost. For the PS and GA methods, increasing the number of trajectories $P$ will significantly decrease the risk of being trapped in a local minimum. At the same time, it will also increase the computational costs of these methods, justifying the need for faster techniques.
\begin{table}[b!]
\begin{center}
\begin{tabular}[h]{|c|c|c|c|c|c|} \hline
&       $r_s^x$         &        $r_s^y$        &  $S_0$  & CPU time (s)& Number of model runs \\ \hline
SA & $0.07015$ &  $0.1710$ &  $1.15\%$ & $3987.1$ & $28,191$ \\ \hline
PS & $0.055$ & $0.1534$ & $1.07 \%$ & $127.3$ & $1,072$ \\ \hline
GA & $0.0447$ & $0.1577$ & $0.98\%$ & $121.8$ & $1,232$ \\ \hline
\end{tabular}
\end{center}
\caption{Performance of global optimization methods. The first three columns describe the errors with respect to true location and intensity of the source. The last two columns present the computational time and number of model runs.}\label{table::global_techniques_output_Exp1}
\end{table}

By coupling the IF method to each of the global techniques, as described in Section \ref{sec:hybrid_techniques}, we will decrease the computational load of the global approaches. This will be accomplished by exploring the fast convergence rate of the IF method and truncating the last stage of the global optimization methods known for low convergence rate. For such hybrid techniques to be successful, we need to properly select a sub-domain $\Omega_0 = [r_s^{x**}-a_1,r_s^{x**}+a_1]\times[r_s^{y**}-a_2,r_s^{y**}+a_2]\times[S_0^{**}-a_3,S_0^{**}+a_3] \subset \Omega$ such that the true source location and intensity  $\pt_0$ is included in $\Omega_0$. Our hope is that in case the sub-domain $\Omega_0$ is small enough; i.e, $a_1~,a_2,$ and $a_3$ are small, so the chance for the objective function to have multiple minima across $\Omega_0$ is decreased and we can exploit the fast convergence rate of IF to identify the source properties. Given the constraints, the question is how large to select $\pmb{a} =[a_1,a_2,a_3]$ with no information about the true radiation source properties.  The answer depends also on $\pt^{**} = [r_s^{x**},r_s^{y**},S_0^{**}],$ the suboptimal point obtained by the truncated global optimization techniques. Unfortunately there is no way to actually measure the distance between the current point and the true properties of the radiation source with no prior information.

Consequently, the decision regarding when to stop the global techniques and how to select $\pmb{a}$ relies on empirical considerations. By selecting $\pmb{a} = [10,10,10^{10}]$, we are able to define sub-domains $\Omega_0$ based on the outputs of global optimization methods. This choice proved to be successful given appropriate stopping early criteria for the global techniques; thus no other choices of $\pmb{a}$ have been tested. Initially we seek to identify one appropriate early stopping criteria by trial and error.

\paragraph{Example 2} First we test our hybrid techniques using an early stopping criterion for the global methods based on restricting the number of model runs. Initially, only the results obtained from one simulation are discussed. Later, the outputs resulted from $10$ different simulations are compared to account for the stochastic nature of the algorithms. Consequently we constrain the global techniques to terminate after $336$ model runs using only $16$ trajectories. The design settings are similar to the ones used in Example $1$, as summarized in Table \ref{tbl::global_param}, except for the reannealing parameter $r_p$ which is set to $10$.

\begin{figure}[b!]
  \centering
  \subfigure[] {\includegraphics[scale=0.35]{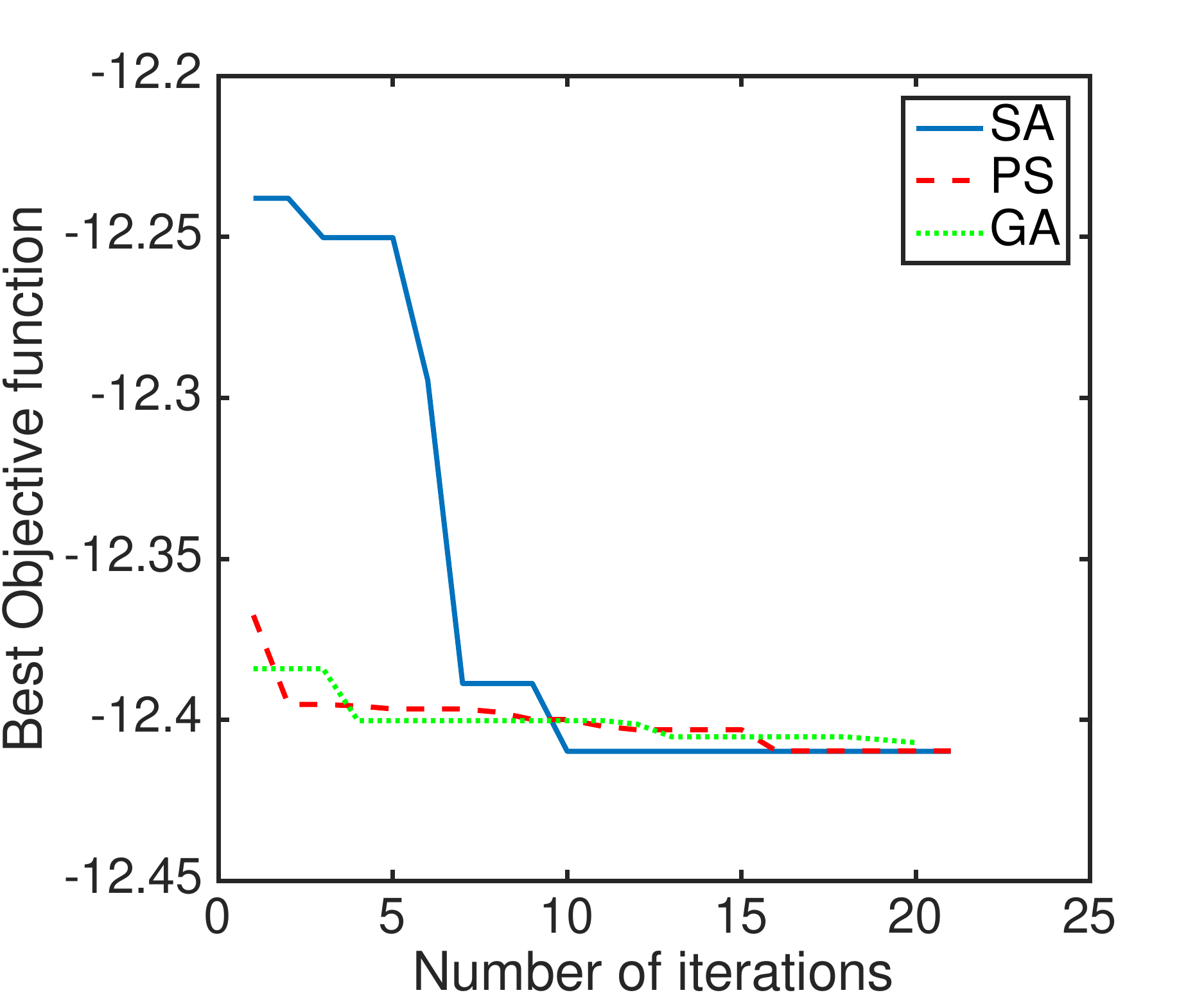}}
  \subfigure[]{\includegraphics[scale=0.35]{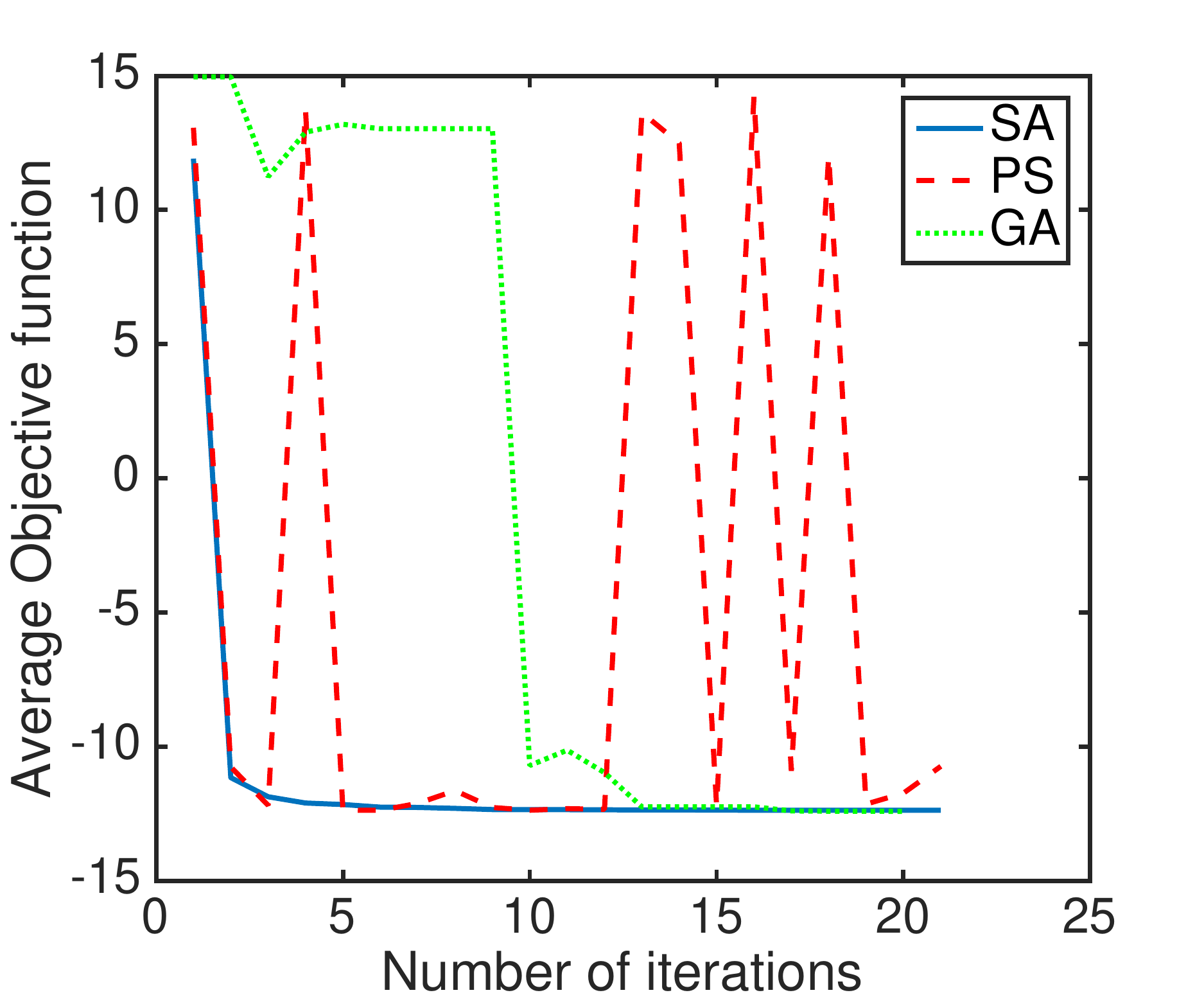}}
\caption{ The performances of global optimization techniques: (a) best objective function; (b) averaged objective function over the entire population of trajectories.}
\label{Fig::Gl_Opt_cost_description_poisson}
\end{figure}

Figure \ref{Fig::Gl_Opt_cost_description_poisson} illustrates the evolution of the global optimization objective functions. In panel (a), the smallest objective function values are plotted and are referred to as best. For the SA method, the displayed values result from a single trajectory corresponding to the smallest objective function among all the threads obtained at the end of the optimization process. Thus, even if an uphill point was selected, the objective function of the previous point is displayed. These flat local regions are easy to see in Figure \ref{Fig::Gl_Opt_cost_description_poisson}.  In contrast, for the PS and GA methods, at each iteration, the plot includes the smallest objective function among all the trajectories.

Information regarding the mean optimization path is provided in Figure \ref{Fig::Gl_Opt_cost_description_poisson}(b) and we can extract some clues regarding the spatial regions explored during optimization. The SA independent trajectories quickly leave the areas near the sensor locations and uphill points do not visibly increase the mean of the objective function values. This is not the case for some of the trajectories of GA, which only depart the sensor neighborhoods in the last part of the optimization process. A different pattern is noticed for PS population which continued to revisit these areas as evidenced by the spikes
in the average objective function.

\begin{table}[b!]
\begin{center}
\begin{tabular}[h]{|c|c|c|c|c|} \hline
        &       $r_s^x$         &        $r_s^y$        &  $S_0$  &    CPU time (s)  \\ \hline
SA & $152.636$ & $99.416$ & $3.906 \cdot 10^9 $ & $29.1$ \\ \hline
PS & $156.627$ & $100.332$ & $4.276 \cdot 10^9$ & $45.7$ \\ \hline
GA &  $151.134$ & $100.781$ & $4.4321 \cdot 10^9$ & $46.5$ \\ \hline
\end{tabular}
\end{center}
\caption{The pseudo estimates $\pt^{**}$ and computational times of the global optimization techniques.}\label{table::global_techniques_output1}
\end{table}

\begin{figure}[b!]
  \centering
  \subfigure[] {\includegraphics[scale=0.28]{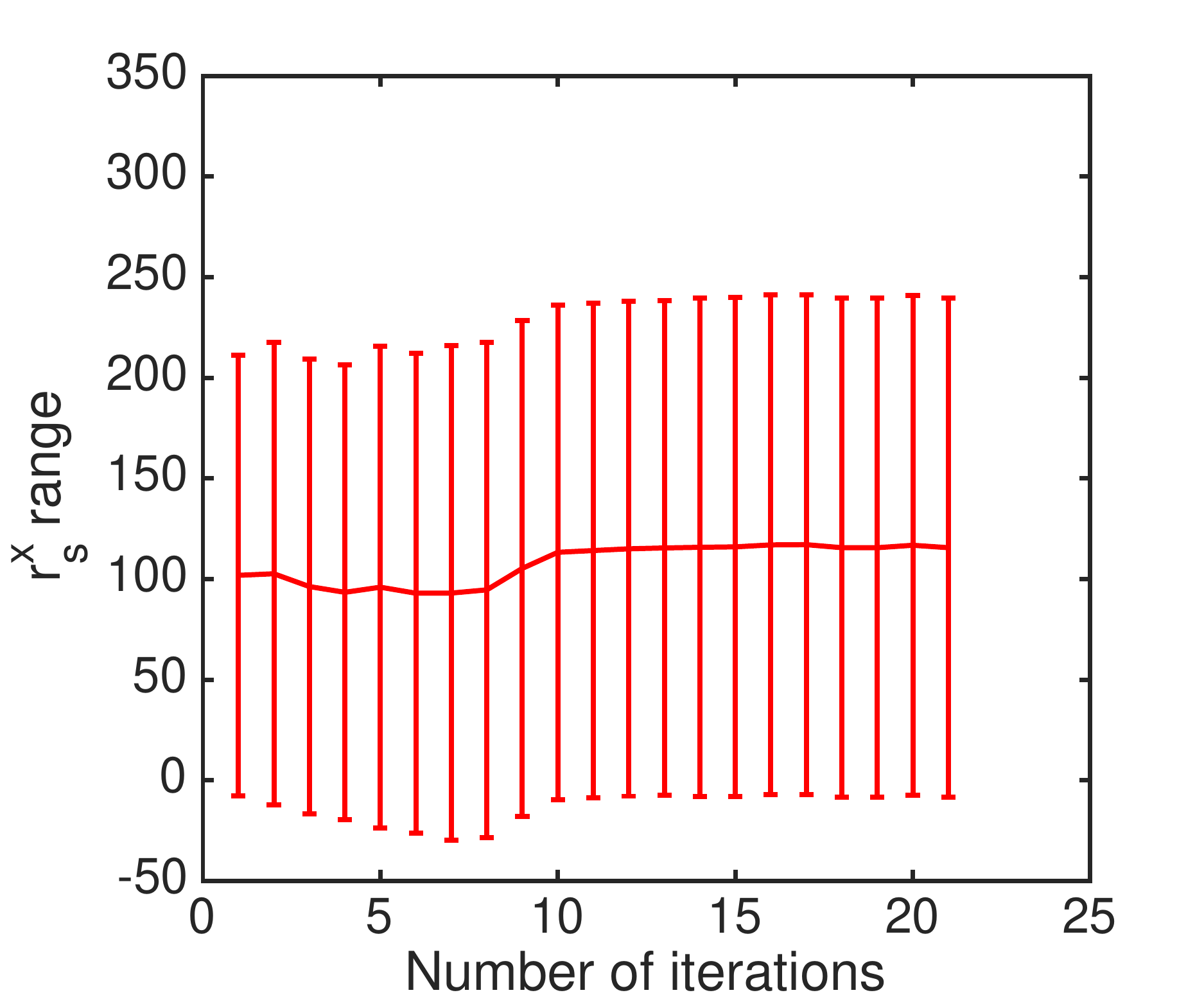}}
  \subfigure[] {\includegraphics[scale=0.28]{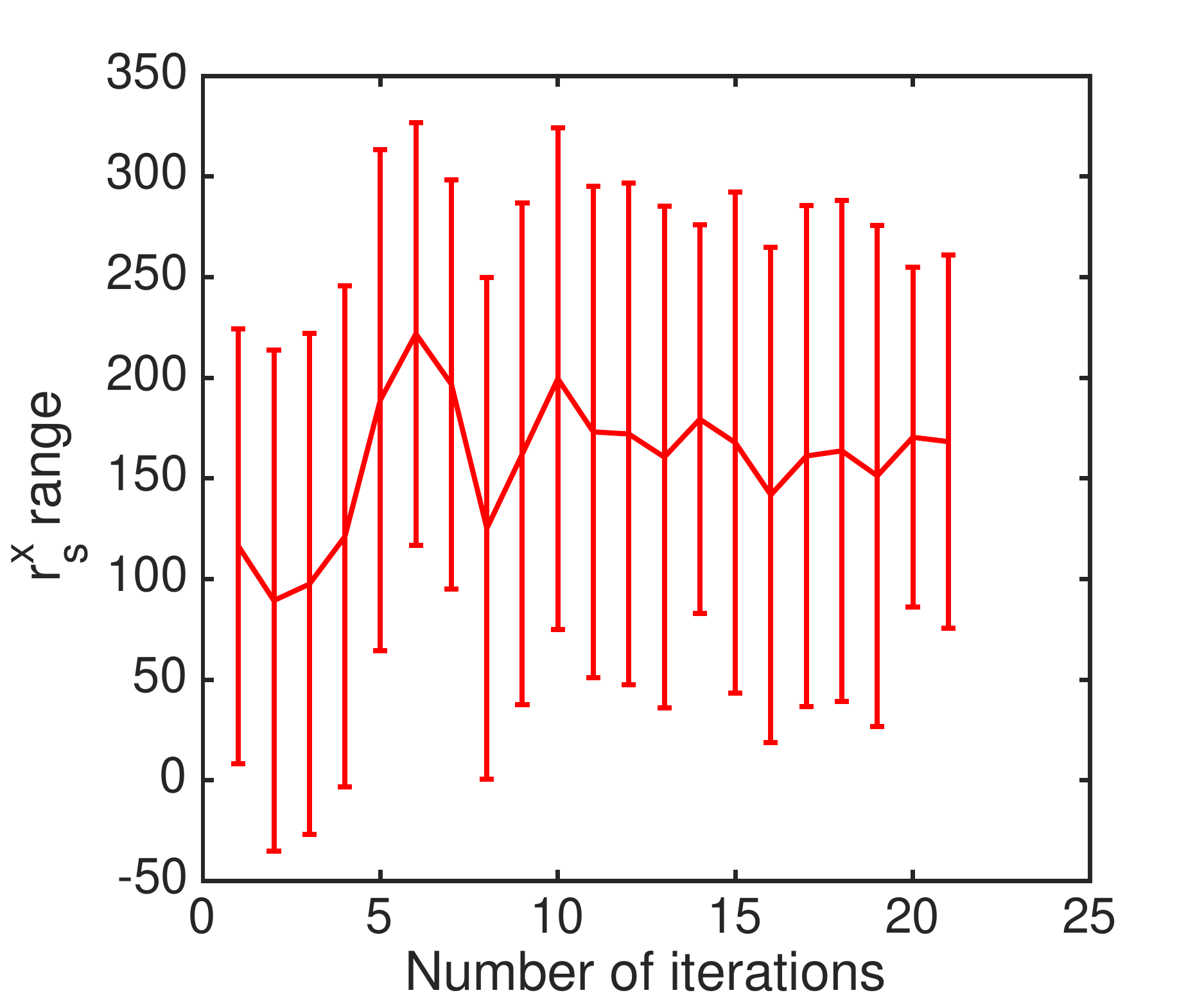}}
  \subfigure[] {\includegraphics[scale=0.28]{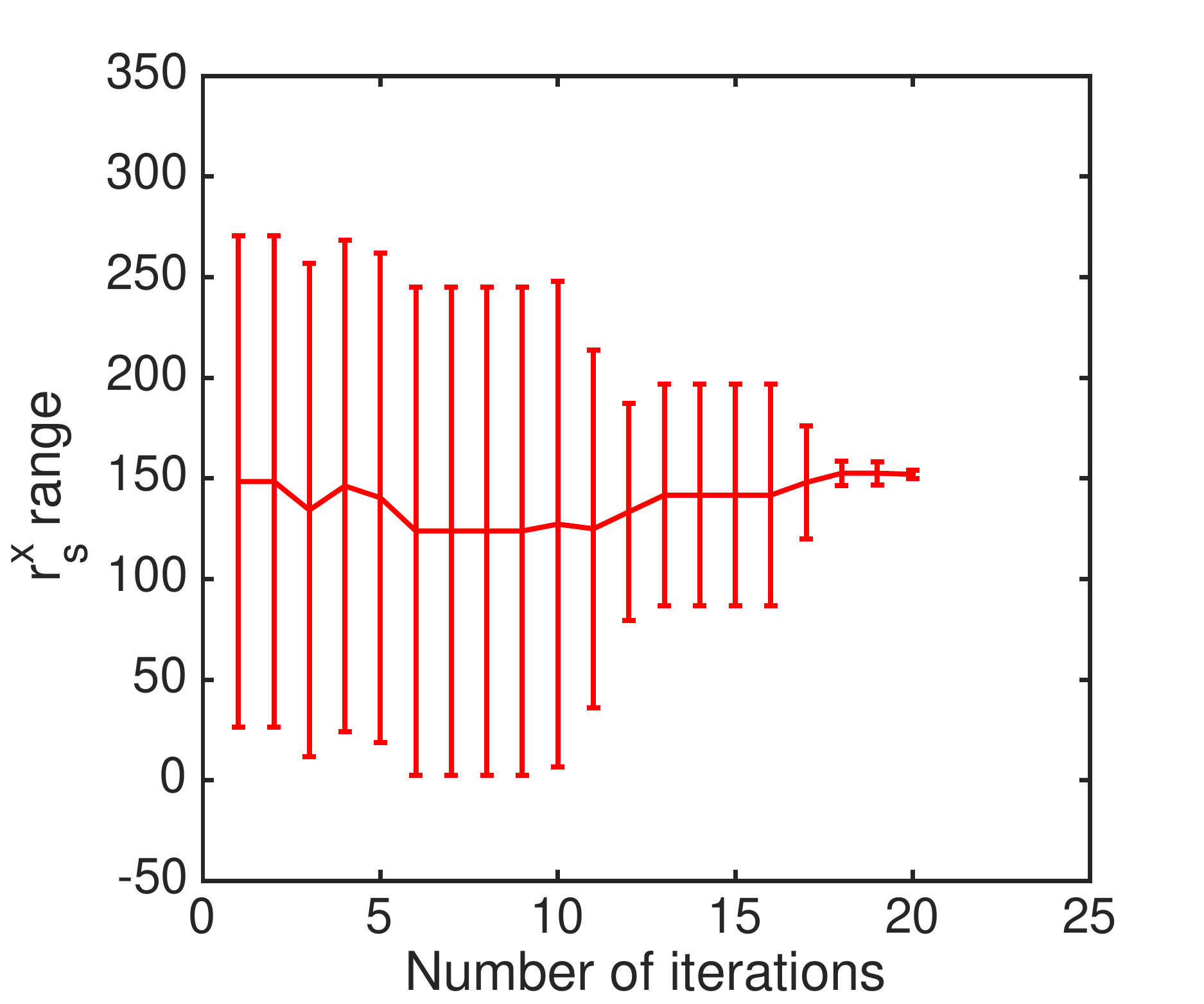}}
\caption{Ranges of searched $r_s^x$ during the global optimization methods runs: (a) SA, (b) PS, (c) GA.}
\label{Fig::Gl_opt_ranges_x}
\end{figure}
\begin{figure}[t!]
  \centering
  \subfigure[] {\includegraphics[scale=0.28]{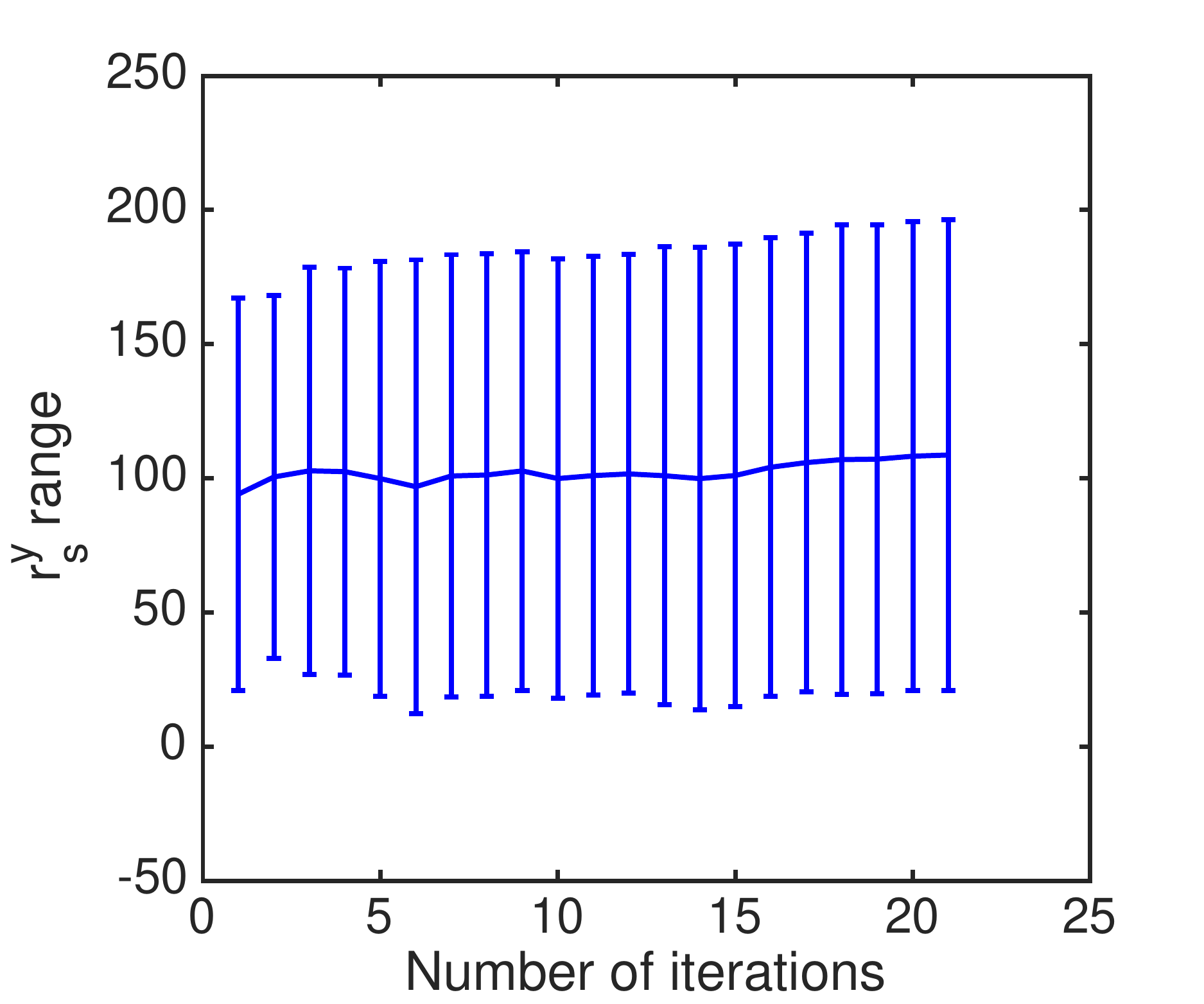}}
  \subfigure[] {\includegraphics[scale=0.28]{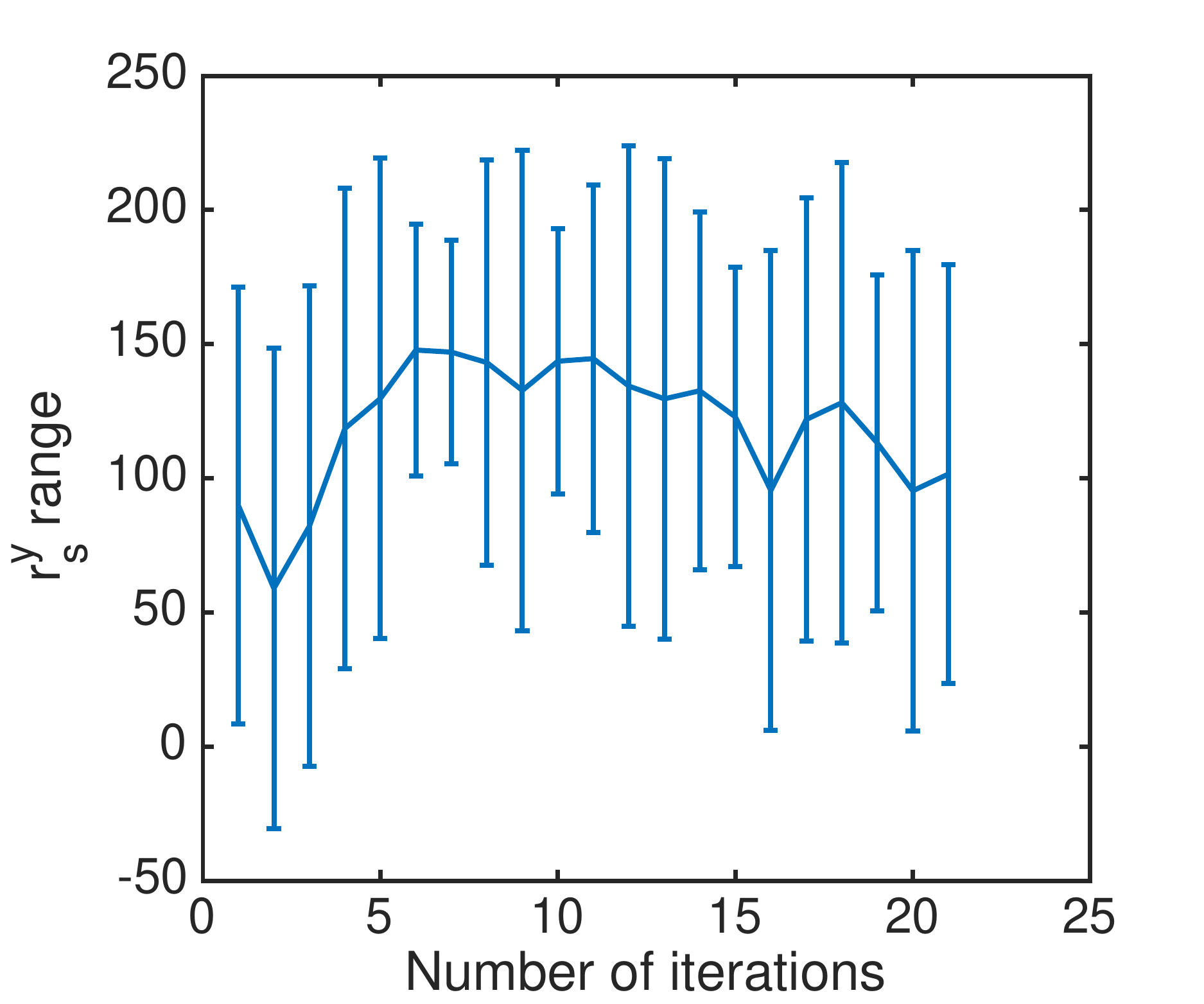}}
  \subfigure[] {\includegraphics[scale=0.28]{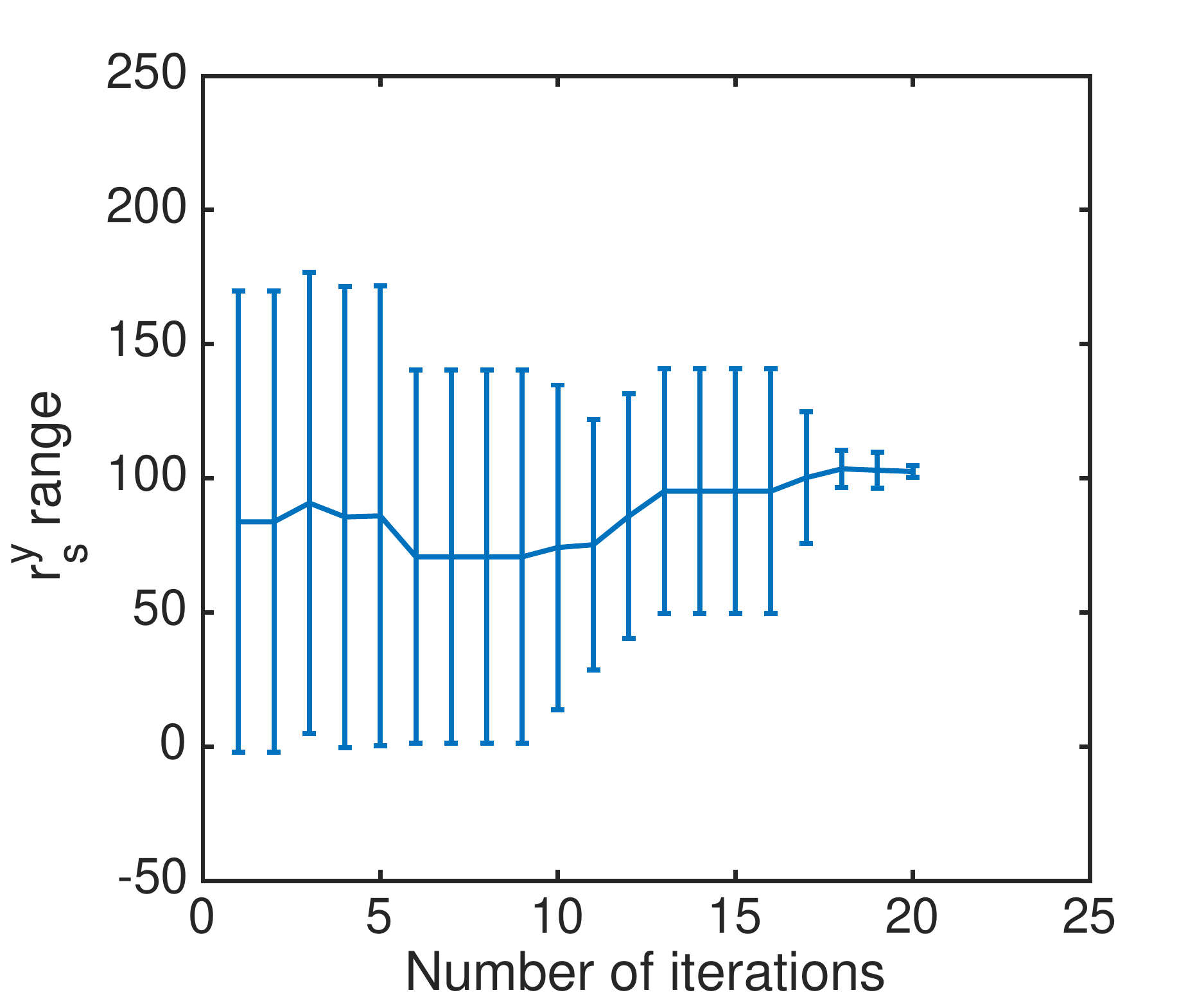}}
\caption{Ranges of searched $r_s^y$ during the global optimization methods runs: (a) SA, (b) PS, (c) GA.}
\label{Fig::Gl_opt_ranges_y}
\end{figure}
\begin{figure}[t!]
  \centering
  \subfigure[] {\includegraphics[scale=0.28]{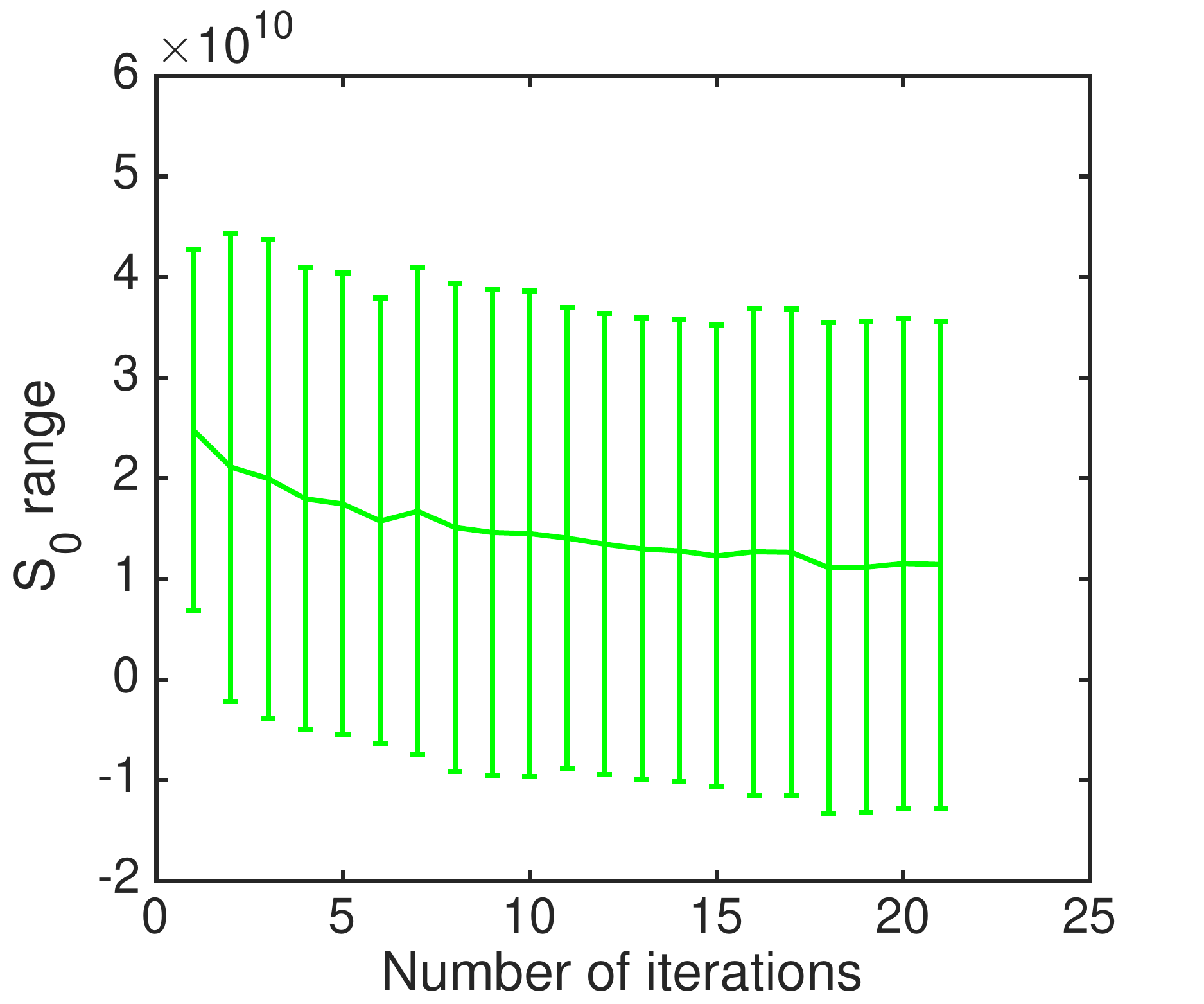}}
  \subfigure[] {\includegraphics[scale=0.28]{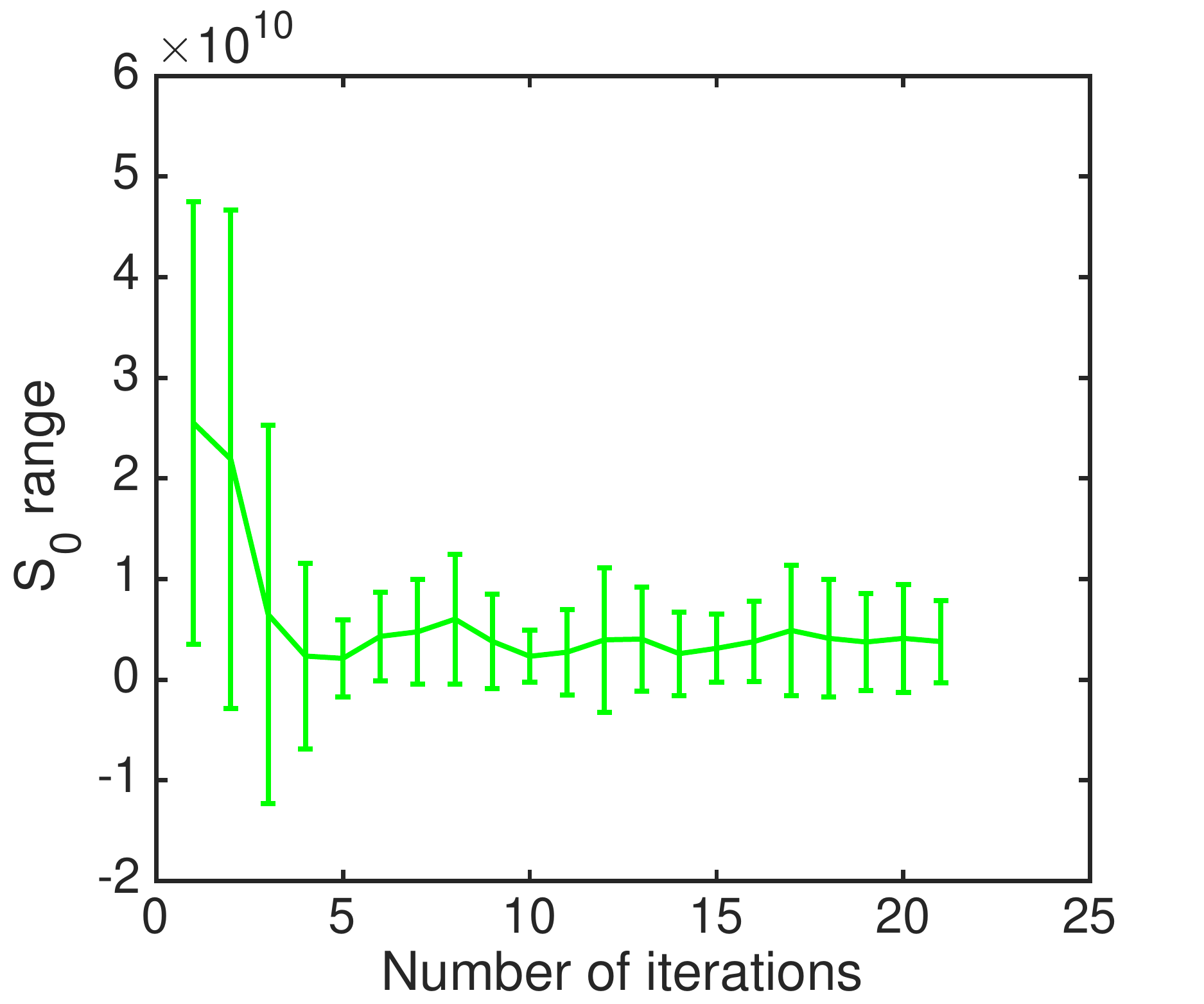}}
  \subfigure[] {\includegraphics[scale=0.28]{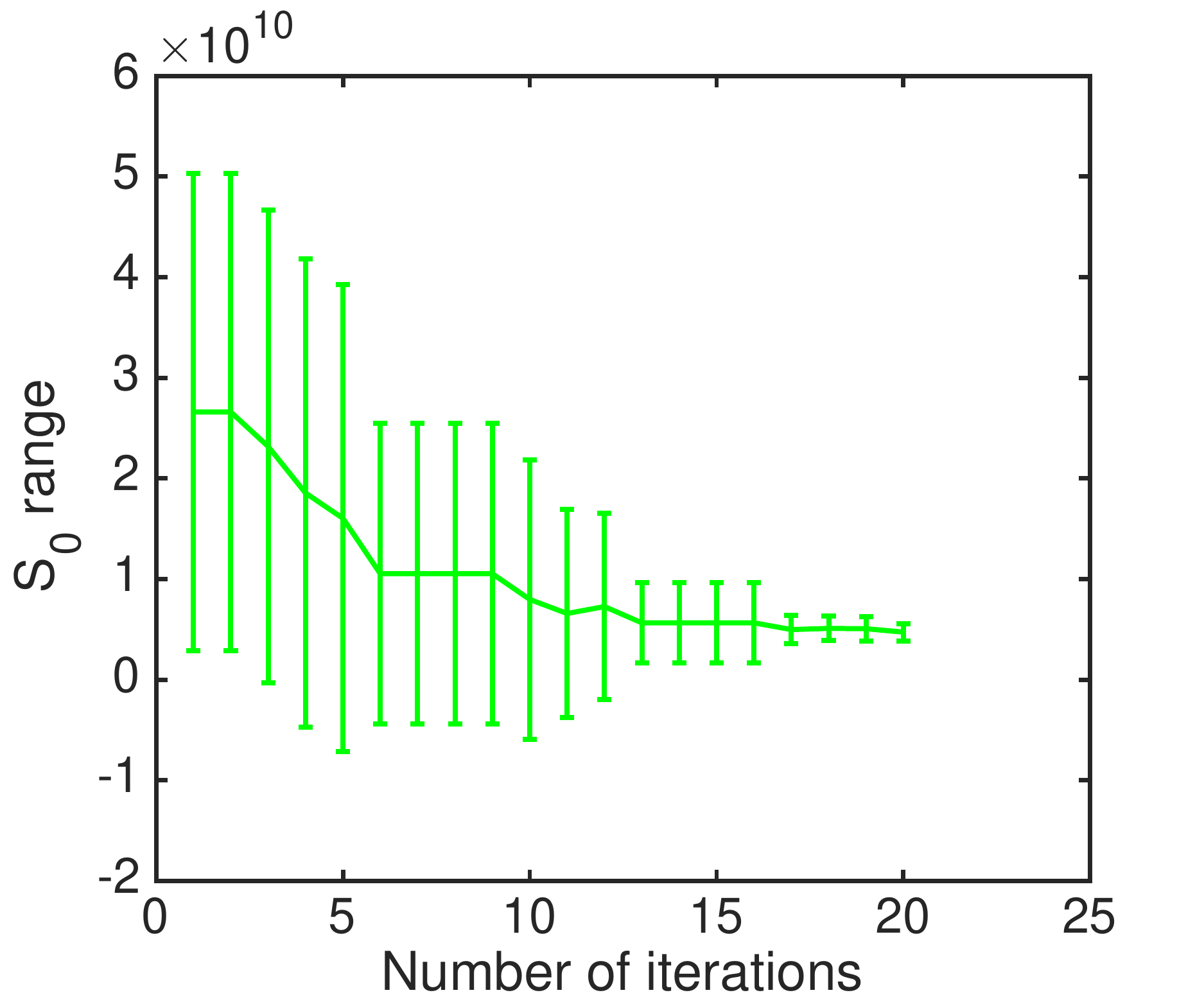}}
\caption{Ranges of searched $S_0$ during the global optimization methods runs: (a) SA, (b) PS, (c) GA.}
\label{Fig::Gl_opt_ranges_S}
\end{figure}

We illustrate in Figures \ref{Fig::Gl_opt_ranges_x}-\ref{Fig::Gl_opt_ranges_S} direct measures of the amount of space variability exhibited by the cloud of points at every iteration for all the global methods. Each property of the source is shown separately. As expected, the larger spread is noticed in the case of SA points.
The distances between the components are larger for the PS as compared with those of GA but are much smaller than in the case of SA. This is explained by the nature of the algorithms. Whereas all of the techniques start their trajectories from uniformly distributed points inside the feasible space, the PS and GA algorithms quickly move their search in the areas corresponding to the larger likelihood function values. At the end of the optimization process, all of the PS and GA trajectories converge to the same area inside the feasible space but not necessarily the global optimum. This does not happen for SA trajectories due to their independent nature, separated threads may end up pointing toward far away regions inside the search space. By employing a sufficient, large number of trajectories, it is almost certain that at least one will hit the space area where the optimum point lies.

The estimated source location and intensity obtained after $336$ model runs are compiled in Table \ref{table::global_techniques_output1}. These results reveal that PS has obtained the most accurate space components of the radiation source. For the same number of model runs, SA has the most efficient parallel implementation finishing the search in $29.1$ s. More importantly, the generated sub-domains $\Omega_0$ secured by all the techniques include the true values of the source components $\pt_0$. This provides the desired initial configuration for the IF method. For this single run test, the a-priori selection of ${\bf a}$ is adequate.

% assuming that the objective function inside this small domain admits only one local minimum.

%
\begin{table}[b!]
   \begin{center}
  \begin{tabular}{| c| }
    \hline
    Implicit Filtering \\ \hline
    $ \textrm{budget} = 300,~\textrm{maxitarm} = 3,~\textrm{maxit} = 50,~\tau = 1.2|J(\pt)\times 10^{-20}|,~\beta = 1,~h_{\min} = 2^{-15},~h = 2^{-1}. $ \\
    \hline
  \end{tabular}
\end{center}
       \caption{Design parameters of local optimization method. See Appendix D for more details.}
       \label{tbl::local_param}
\end{table}

The design parameters for the IF method are summarized in Table \ref{tbl::local_param}.
The settings include a budget of no more than $300$ model runs, a maximum number of $3$ step size reductions within the modified line search, the initial line search stepsize $\beta=1$ and the maximum number of  Quasi-Newton directions for the same stencil size $h$ is set to $50$. The initial choice for $h$ is $2^{-1}$, whereas the smallest value allowed for the stencil size is set to $h_{\min} = 2^{-15}$. One of the quasi-Newton iteration stopping criteria depends on the initial entering value $\pt$, i.e. $\tau = 1.2 |J(\pt)|\cdot 10^{-20}$.

\begin{figure}[t!]
  \centering
  \subfigure[] {\includegraphics[scale=0.35]{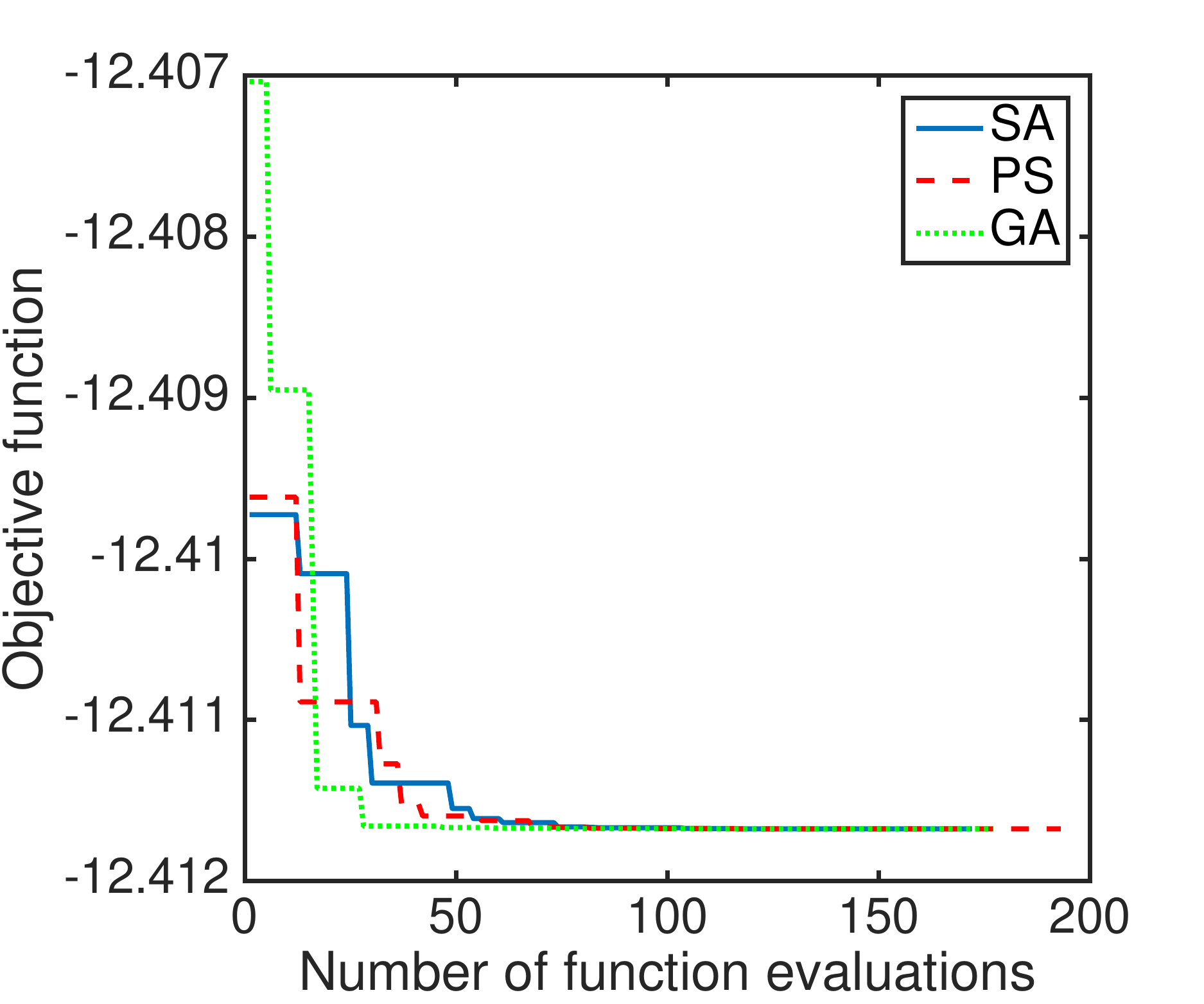}}
  \subfigure[] {\includegraphics[scale=0.35]{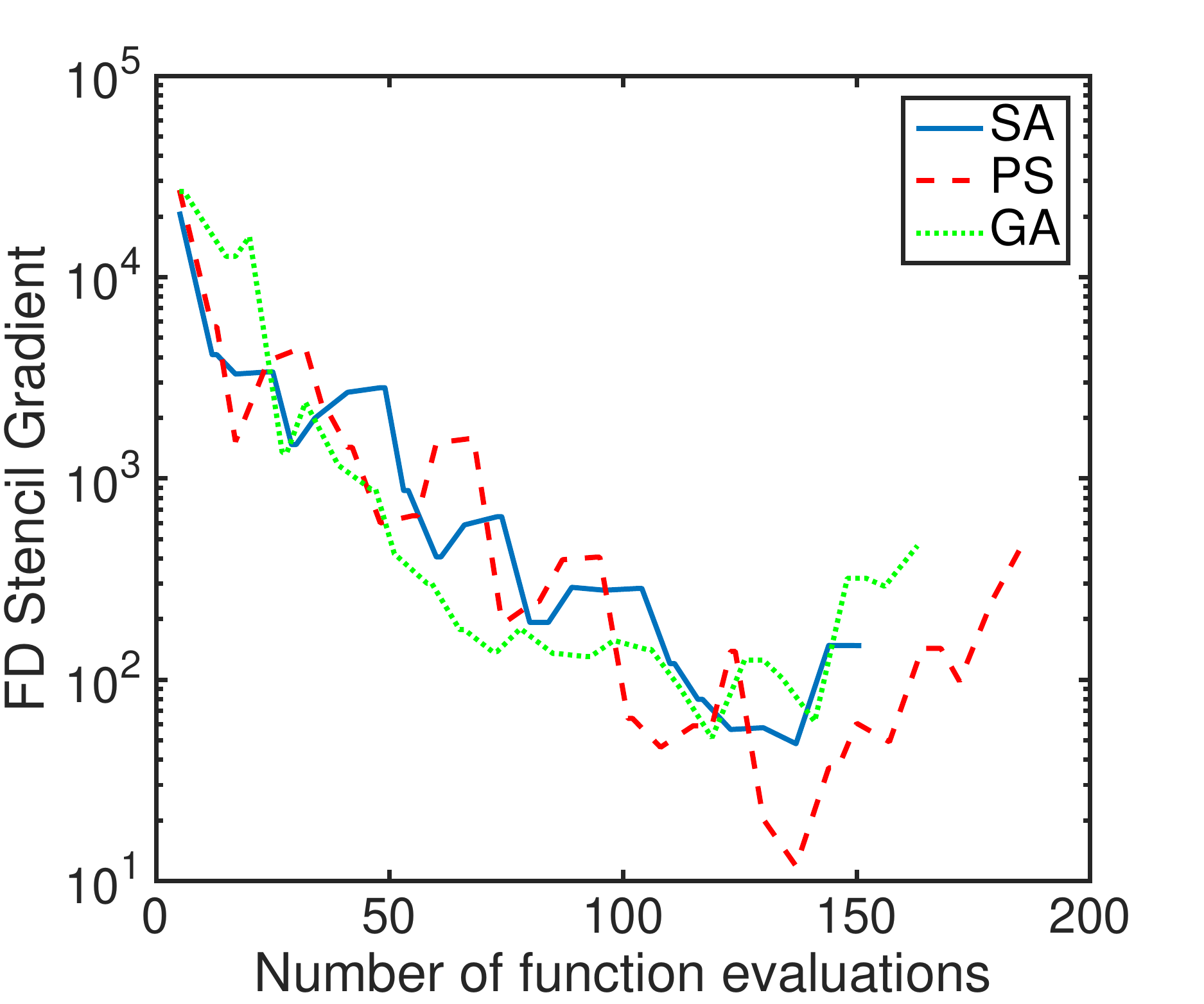}}
  \caption{Implicit Filtering performances for different starting points corresponding to the outputs of Simulating Annealing, Particle Swarm and Genetic Algorithm: (a) cost function and (b) FD stencil gradient.}
 \label{Fig::SA_PS_GS_IF_poisson}
\end{figure}

The objective function and its finite difference estimate of the stencil gradient are plotted in Figure \ref{Fig::SA_PS_GS_IF_poisson}. These were obtained for different IF simulations having various starting points obtained via global optimization techniques. The decrease in the objective function is smaller than in the case of a global methods as seen in Figure \ref{Fig::Gl_Opt_cost_description_poisson}. This suggests that the local optimization component is less computational demanding. This is also underlined by the number of model runs showed in Table \ref{table::hybrid_techniques_output}.

For a continuous differentiable objective function, the norm of the gradient decreases as one approaches an optimum point. We notice a different behaviour in Figure \ref{Fig::SA_PS_GS_IF_poisson}b due to the lack of smoothness. Whereas the gradient of the objective function may not be defined due to the discontinuities of the likelihood function, finite-differences can always be computed. The IF method makes use of this type of information to compute the quasi-Newton direction. At the same time, the line search is modified so that a simple decrease in the objective function is sufficient; i.e., the local method does not trust that the finite differences could represent the gradient. We employ the negative log of the absolute values to display the results in the left panel while logarithmic scale is used to show the norm of the finite difference estimates in the right panel. Starting from different outputs of the discussed global optimization methods leads to mildly different computational costs.  More model runs are required by the IF to finish the search when started with the output produced by the PS algorithm. Initiating the local method with the SA output requires the smallest number of model runs to achieve convergence.   All  simulations are terminated once the list of scales for $h = $ $\left[2^{-1},2^{-2},..,2^{-15}\right]$ has been exhausted.

\begin{figure}[t!]
  \centering
   \subfigure[] {\includegraphics[scale=0.28]{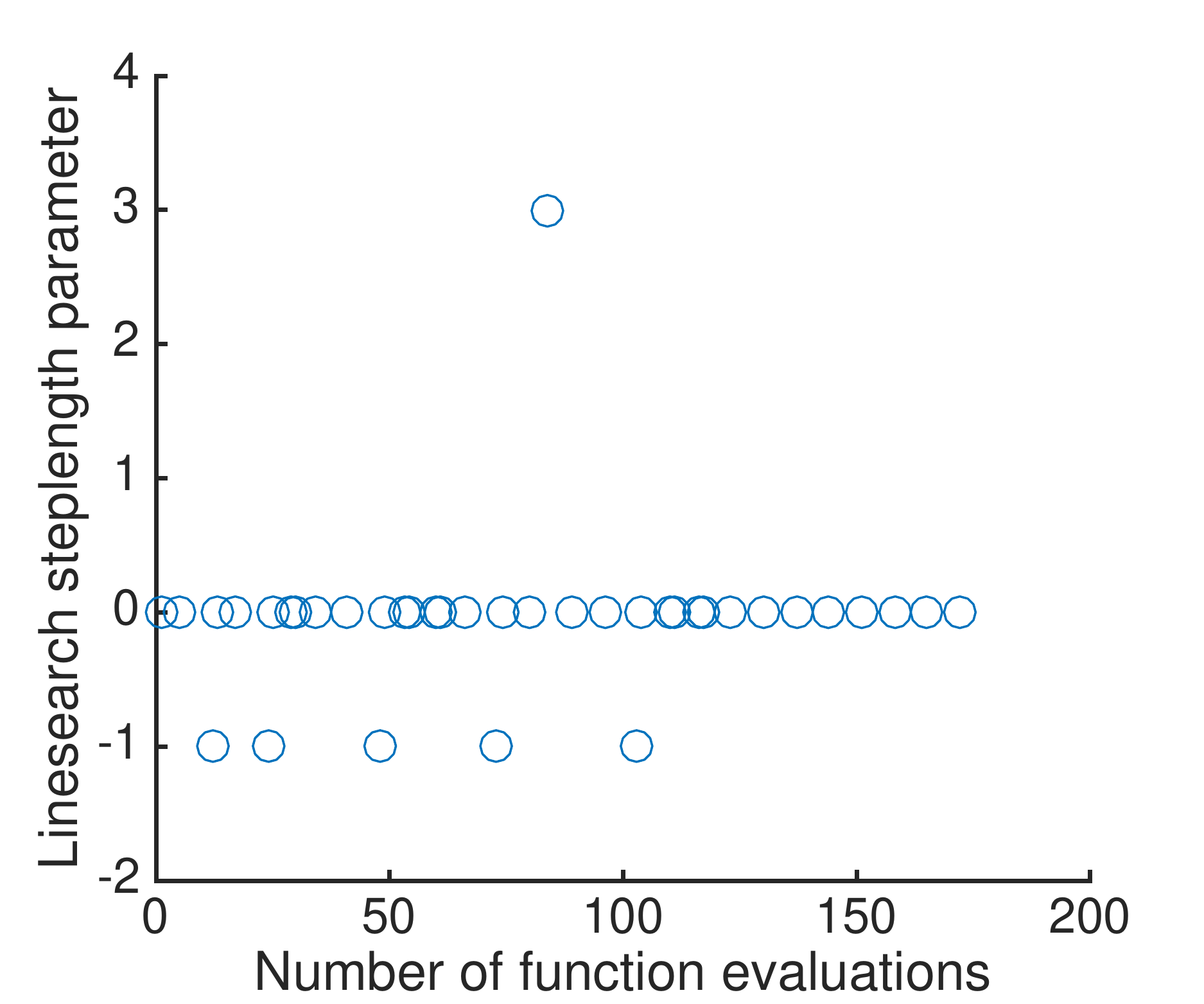}}
  \subfigure[] {\includegraphics[scale=0.28]{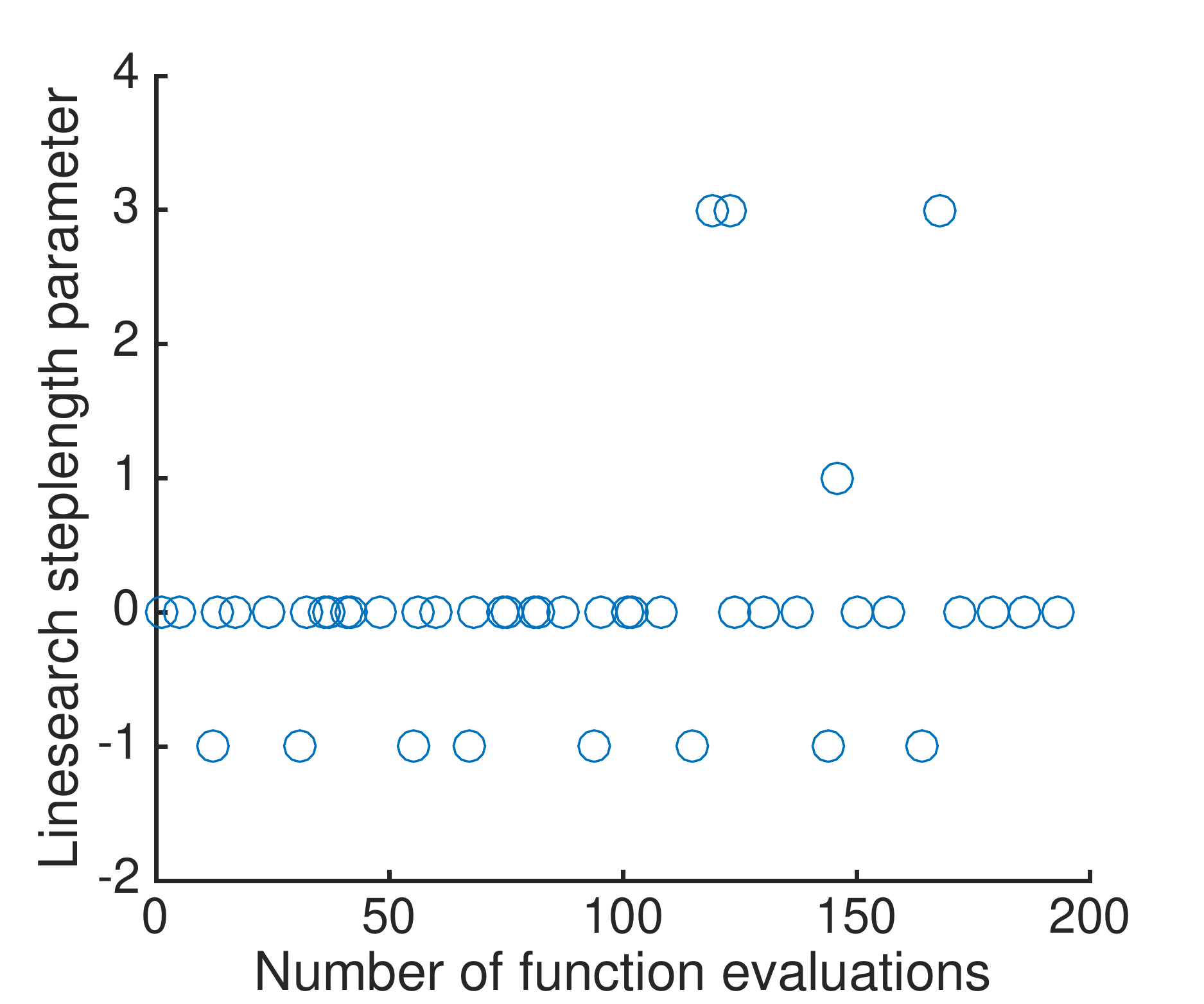}}
   \subfigure[] {\includegraphics[scale=0.28]{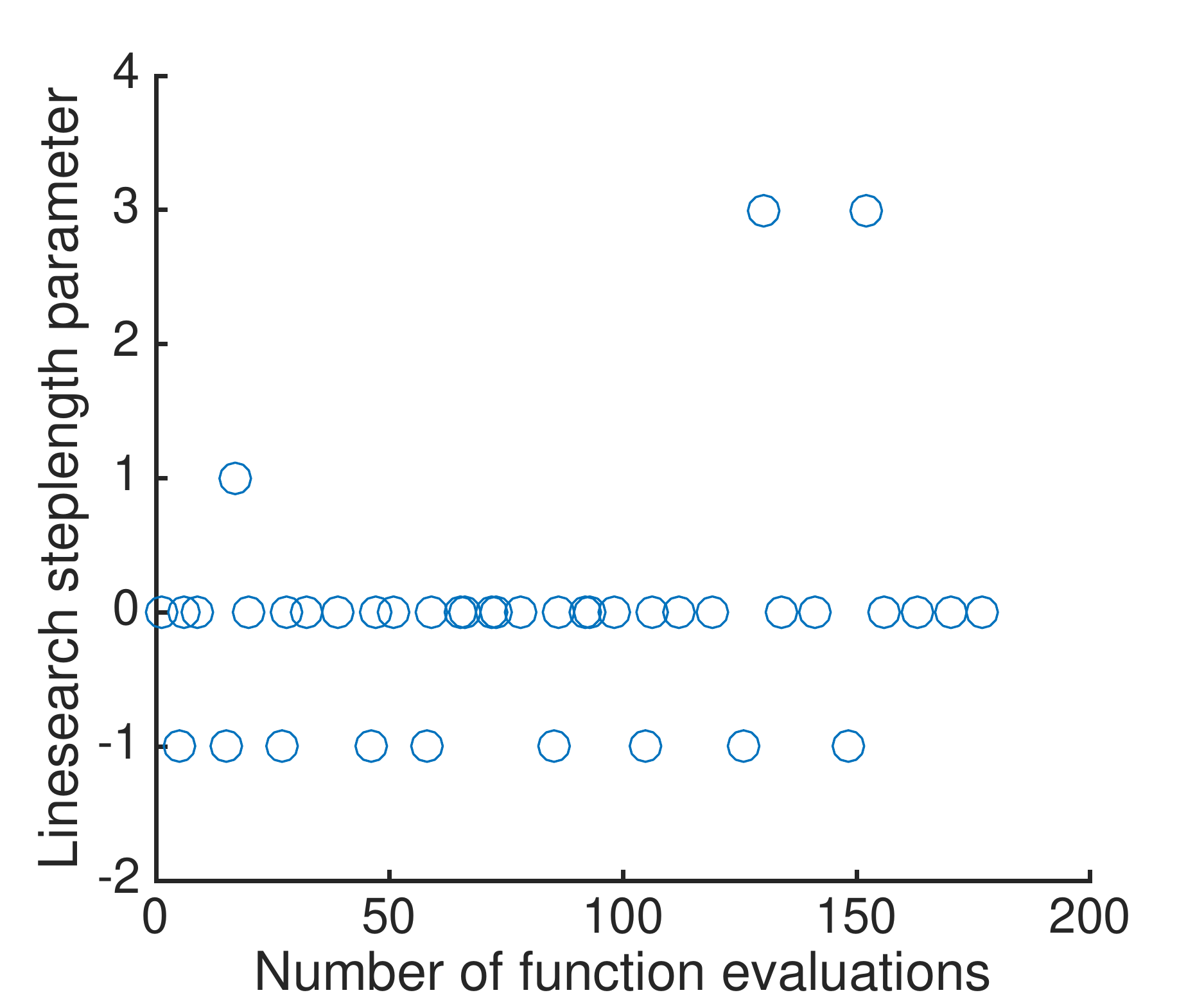}}
\caption{Line search parameters of the Implicit Filtering method for different initial starting points provided by: (a) Simulated Annealing, (b) Particle Swarm and (c) Genetic Algorithm.}
\label{Fig::Linesearch_IF_poisson}
\end{figure}

At the end of the search, the results in Figure \ref{Fig::SA_PS_GS_IF_poisson}(b) reveal an actual increase in the norm of the finite difference estimation of the stencil gradients. This could be explained by the small magnitude of the scale $h$. Figure \ref{Fig::Linesearch_IF_poisson} describes the number of times the step length was reduced during the line search. A value of $-1$ shows that the algorithms encountered a stencil failure in the previous iteration, while a value of $3$ indicates the line search failed, but the stencil poll found better points on the stencil. For the majority of times, no stepsize reductions were required to generate downhill points during the modified line search stage. The errors in the estimates of the radiation source properties together with the computational times and numbers of model runs for all the hybrid techniques are shown in Table \ref{table::hybrid_techniques_output}. The errors for the source intensity are relative.
\begin{table}[t!]
\begin{center}
\begin{tabular}[h]{|c|c|c|c|c|c|} \hline
        &       $r_s^{x0}$         &        $r_s^{y0}$        & $S_0^0$   & CPU time (s) & Number of model runs   \\ \hline
SA+IF & $0.009$ &  $0.194$ & $1.49\% $ & $95.2$ & $508$ \\ \hline
PS+IF &$0.056$ & $0.179$  & $1.04\%$ &  $126.8$ & $529$ \\ \hline
GA+IF & $0.002$ & $0.1662$ & $1.28\%$ &  $121.1$  & $513$ \\ \hline
\end{tabular}
\end{center}
\caption{Errors of the estimates obtained by the hybrid algorithms together with their computational cost performances.}\label{table::hybrid_techniques_output}
\end{table}

As seen from the overall number of function evaluations one may conclude that the computational complexities of all the hybrid techniques are similar. However, it is important to mention that a general stopping criterion based on the number of model runs was imposed for all the techniques in the first stage. Thus the differences in function evaluations are obtained only from the IF stage.

To take into account the stochastic nature of the global techniques, $10$ simulations have been performed. Same configurations have been utilized and the resulting sub-domains have been verified. Out of $10$ runs, only in $5,~3$ and $2$ simulations performed by SA, PS and GA, respectively, the proposed sub-domains contain the true location and intensity of the source. This suggests that more function evaluations are required in order to obtain, for every run, sub-domains $\Omega_0$ having the global optimum as interior point. This result also indicates that SA can reach the proximity of the global optimum more often than PS and GA at the present design configurations.

%Of course other strategies could be pursued such as increasing the value of $\pmb{a}$. We did not followed this route in this paper and we decide to change the design parameters configurations of the methods and maintain the size of the sub-domain $\Omega_0$.

\paragraph{Example 3} We test our hybrid methods using an early stopping criterion for the global methods based on a target objective function value. Following this approach, we are able to identify the number of model runs required by each global method to generate sub-domains $\Omega_0$ that include the true radiation source properties. We will compare, in an average sense, the coupled global and local optimization techniques against the solely global methods and DRAM and DREAM algorithms. We focus on the accuracy of the estimates and the computational cost of the methods.

We set the objective function target value to $-2.452\times 10^5$ informed by the previous simulations. We also relaxed the previous stopping criterion to allow  a maximum of $3000$ function evaluations. The design configurations are described in Table \ref{tbl::global_param_config2}. Here we select the number of trajectories $P = 70$ for all the methods. A larger number of trajectories leads to a better space exploration with higher computational costs. The GA configuration summarized in Table \ref{tbl::global_param} is generic and depends on the number of trajectories.
\begin{table}[b!]
\begin{center}
  \begin{tabular}{ |c | c |} \hline
    SA & PS \\ \hline
    $T_1^0 = T_2^0,~T_3^0 = 100,~$ & $\textrm{Ns} = 17,~W = 1.1$    \\ \hline
    $r_p = 30$ & $y_1 = y_2 = 1.49$  \\
    \hline
  \end{tabular}
  \end{center}
     \caption{Design parameters of global optimization methods. See Appendix A and B for more details.}
     \label{tbl::global_param_config2}
\end{table}
Here $10$ simulations were performed and the results were averaged to account for the stochastic nature of the global optimization techniques. For all simulations, all of the methods were able to decrease the objective function below the predefined target with generated sub-domains $\Omega_0$ that include the true source location and intensity. The results are provided in Table \ref{table::hybrid_techniques_output4}. Whereas the SA implementation exhibits the smallest overhead time, the PS method requires the smallest number of function evaluations, which is more than three times lower than the number of function evaluations needed by the SA method. Moreover, for a sufficiently large number of trajectories, communication between chains enhances the search performance, as seen from the number of function evaluations corresponding to PS and GA. A slight decrease in the accuracy of the GA estimate can be also observed.
\begin{table}[b!]
\begin{center}
\begin{tabular}[h]{|c|c|c|c|c|c|} \hline
        &       $r_s^{x0}$          &     $r_s^{y0}$ & $S_0^0$ & Number of model runs &  CPU time (s) \\ \hline
SA &    $0.487$   & $0.2672$   & $8.91\%$ &    $4,249$     &     $83.6$     \\ \hline
PS &    $1.097$  &     $0.77$ & $14.35\%$  &     $1,162$     &     $126.9$      \\ \hline
GA & $1.767$  &  $1.095$ & $24.14\%$ & $1,295$       &  $140.8$           \\ \hline
\end{tabular}
\end{center}
\caption{Average performance of global optimization techniques using $10$ simulations.}\label{table::hybrid_techniques_output4}
\end{table}
Next, the ten pseudo-estimates of each of the global techniques are used to start the IF search constrained by the smaller sub-domain $\Omega_0$. The overall computational complexity and accuracy of the hybrid methods are compared in Table \ref{tbl::Overall_results} with the outputs of the global methods. These results are verified against the estimates of the source location and intensity obtained by directly sampling their posterior distributions using DRAM and DREAM.

%Numerical results from DREAM using the Poisson likelihood \eqref{eqn:Poisson_Likelihood}.  The reported parameter estimates are the mean values from the final 25\% of the pooled chain samples.

%
\begin{table}[t!]
\begin{center}
\begin{tabular}[h]{|c|c|c|c|c|c|c|c|c|}
\hline
		& SA+IF & SA & PS + IF & PS & GA + IF & GA & DRAM & DREAM  \\ \hline
No of Func. Eval	& $4,414$ & $139,720$ & $1332.2$ & $4,200$ & $1468.9$ & $5,110$ & $10,000$ & $100,000$	\\ \hline
CPU time (s) &  $149.4$ & $2482.8$ & $197.8$ & $427.2$ & $207.2$ & $521.8$ & $8646.7$ & $37946.1$  \\ \hline
Error $r_s^{x0}$ & $0.083$ & $0.073$ & $0.043$ & $0.072$ & $0.059$ & $0.047$ & $0.06$ & $0.05$\\ \hline
Error $r_s^{y0}$ & $0.197$ & $0.182$ & $0.181$ & $0.1755$ & $0.212$ & $0.1895$ & $0.188$ & $0.180$\\ \hline
Error $S_0^0$ & $1.11\%$ & $1.17\%$ & $1.18\%$ & $1.15\%$ & $0.94\%$ & $0.94\%$ & $1.08\%$ & $1.15\%$\\ \hline
\hline
\end{tabular}
\end{center}
\caption{Numerical results comparing the performances of the hybrid, global and uncertainty quantification techniques. For the hybrid and global techniques the results are obtained by averaging the output of $10$ simulations.}
\label{tbl::Overall_results}
\end{table}

 As expected, the techniques based on connected trajectories exhibit a reduced number of objective function evaluations. The hybrid algorithm combining the PS and IF method evaluated the cost function only $1332.2$ times, which was the smallest among all the hybrid methods. With its independent trajectories, the hybrid SA and IF method proved to have the most efficient parallel implementation, which translated into the smallest computational expense.
 %The communication between the trajectories resumes only in identifying the point in the current iteration associated with the smallest cost  function. This allows for efficient parallelization.
 By combining global and local techniques, we designed fast and accurate hybrid methods which, for this configuration, proved to be $17 \times,~2.1 \times $ and $2.5 \times$ times faster than the global optimization methods. This justifies the effort of employing the IF search method which significantly decreased the numbers of model runs. Overall the hybrid SA+IF, PS+IF and GA+IF, respectively, reduced by $31 \times,~3.1 \times$ and $3.4 \times$ times the number of function evaluations. The minor differences between the error estimates and the true source location and intensity for all the global methods are insignificant due to their stochastic nature.

 DRAM and DREAM were also employed to estimate the most likely location and intensity of the source. By sampling from the posterior distribution, we were able to obtain the marginal densities of the source properties and their associated means. To compare with the hybrid methods, the performances of DRAM and DREAM are included in Table \ref{tbl::Overall_results}. Whereas the Bayesian techniques are far more computationally intensive, they have the significant advantages that they provide posterior input densities detailed in the next section. Coupling reduced order methods and Bayesian methods decreases the computational effort as seen in \cite{attia2016reduced}.

% Dram And Dream Estimates Validate The Results Obtained By The Hybrid And Global Techniques. Some Of The Estimate Errors In The Case Of Uncertainty Quantification Methods Are Slightly Elevated When Compared With The Outputs Of The Hybrid And Global Methods. With So Many Function Evaluations There Is No Doubt That The Estimates Of Dream And Dram Are Accurate With Respect To The Structure Of The Poisson Likelihood. This Is Explained By The Random Effect Existing In The Radiation Behaviour (Observations Errors)  And The Model Discrepancy Used In The Definition Of The Employed Statistical Model \Eqref{Victoras}.

%\subsection{Katie numerical experiments - DRAM}

%The $x$ and $y$ coordinates were bound based upon the limits of the geometry.  The $I_0$ parameter was bound by the interval $[5 \times 10^8, 5 \times 10^{10}]$.
\subsection{Bayesian inference} \label{sec::Bayesian_exp}

We first applied the Delayed Rejection Adaptive Metropolis (DRAM) algorithm detailed in Appendix \ref{sec:Appendix_DRAM} to estimate the radiation source properties $r_s^x$, $r_s^y$, and $S_0$ using the same synthetic data used by the hybrid methods. We employed the Poisson likelihood \eqref{eqn:Poisson_Likelihood} to construct posterior distributions of the source spatial coordinates and intensity.  We used the ordinary least squares estimates obtained using the Nelder–-Mead algorithm \cite{nelder1965simplex} as the starting values for each of the source property chains. For all three source components, we utilized uniform priors constrained by the bounds of the feasible space $\Omega$.

%Figure \ref{Densities} shows the resulting marginal densities.

\begin{figure}[b!]
  \centering
  \subfigure[] {\includegraphics[scale=0.26]{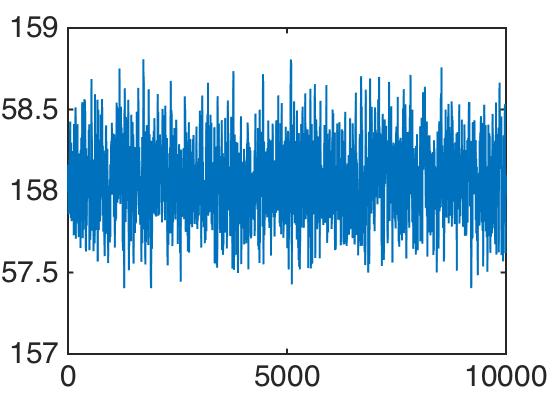}}
  \subfigure[]{\includegraphics[scale=0.26]{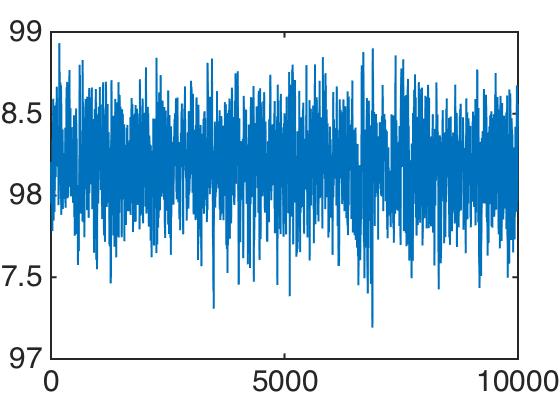}}
  \subfigure[]{\raisebox{1.8mm}{\includegraphics[scale=0.26]{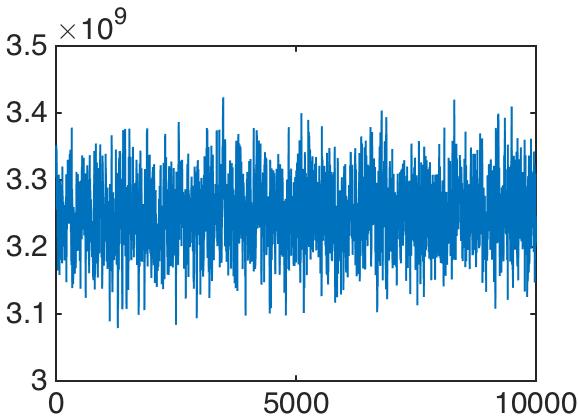}}}
\caption{Chains generated by DRAM for source properties (a) $r_s^x$, (b) $r_s^y$, and (c) $S_0$.}
\label{Fig::Densities}
\end{figure}

After a burn-in period of $3000$ chain iterations, we re-ran the code for $10^4$ iterations.  The resulting chains are shown in Figure \ref{Fig::Densities}.  Visual inspection of the chains shows good mixing and suggests that the chains have converged to the posterior distributions. This is also confirmed by the Geweke diagnostic values, which are $0.99962,~0.99953,~0.99966$ for  $r_s^x,~r_s^y$ and $S_0^0$, respectively.  Using the mean chain values as estimates of our source properties, we obtain $\hat{{\bf r}}_s^0 = (158.06,98.188)$, and $\hat{S}_0^0 = 3.249 \times 10^9$, which compare favorably with the source properties' values used to generate the synthetic data; i.e., ${{\bf r}}_s^0 =(158,98)$, and $S_0^0 = 3.214 \times 10^9$.

\begin{figure}[t!]
  \centering
  \subfigure[] {\includegraphics[scale=0.29]{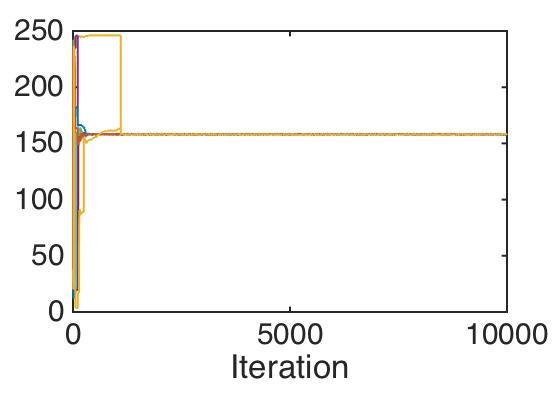}}
  \hspace{-0.1truein}
  \subfigure[]{\includegraphics[scale=0.29]{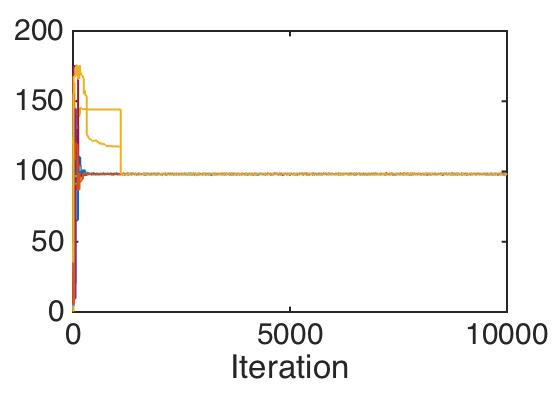}}
  \hspace{-0.1truein}
  \subfigure[]{\raisebox{4.2mm}{\includegraphics[scale=0.27]{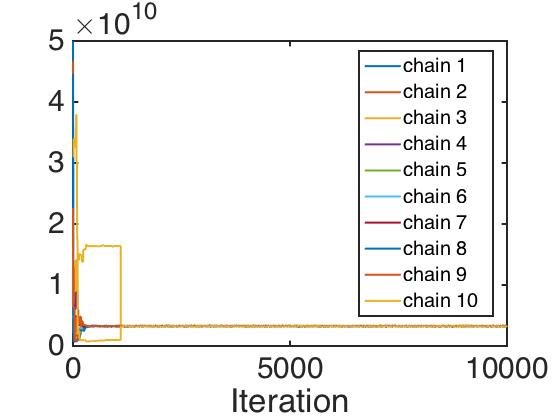}}}
  \subfigure[] {\includegraphics[scale=0.28]{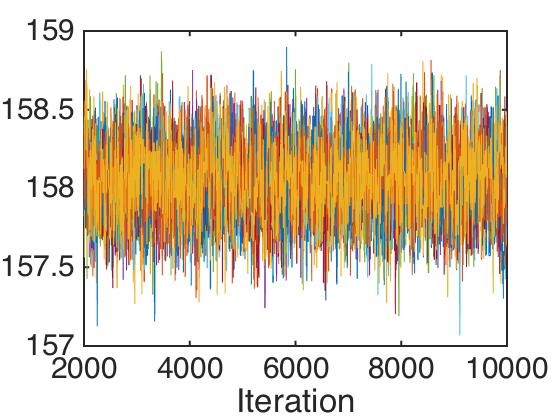}}
  \hspace{-0.1truein}
  \subfigure[]{\includegraphics[scale=0.28]{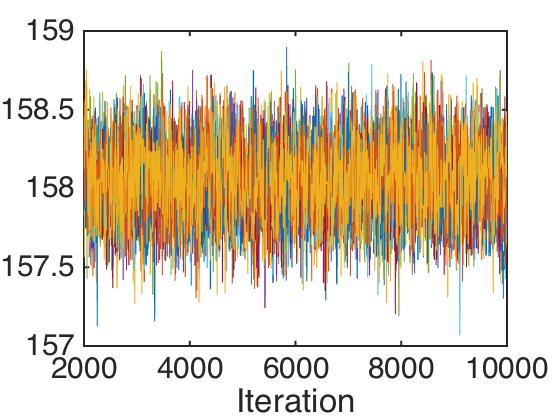}}
  \hspace{-0.1truein}
  \subfigure[]{\includegraphics[scale=0.29]{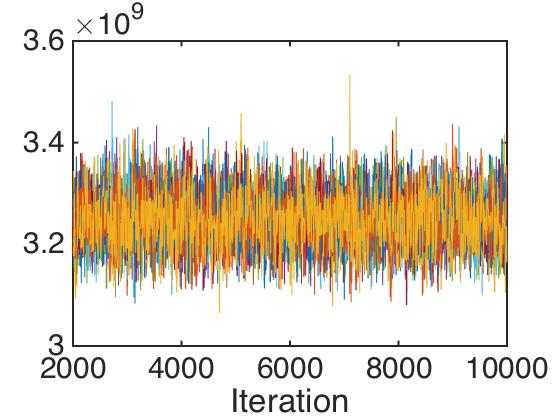}}
\caption{Full DREAM chains for (a) $r_s^x$, (b) $r_s^y$, and (c) $S_0$. Truncated DREAM chains including only the burned-in portion for (d) $r_s^x$, (e) $r_s^y$, and (f) $S_0$.}
\label{DREAMChains}
\end{figure}

To verify our results from Bayesian inference via DRAM, we also estimate the source location and intensity using the DREAM algorithm and the same likelihood and priors. We employ $10$ chains of length $10^4$ for each source component, utilizing a total of $10^5$ function evaluations.  The starting values for each of the $10$ chains are drawn from the same uniform prior distributions; i.e., $U(0,250),~U(0,180)$ and $U(5\times 10^8,5 \times 10^{10})$.

\begin{figure}[t!]
\begin{center}
\includegraphics[width = 3.0in]{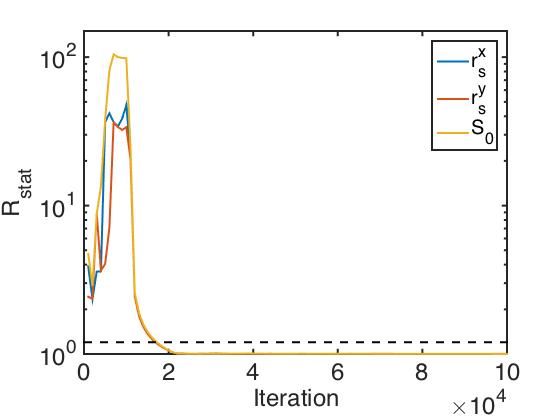}
\end{center}
\vspace{-0.2truein}
	\caption{Plot of the Gelman-Rubin R-statistic at each DREAM chain iteration.  R-statistic values below 1.2 suggest that the chain has converged to its stationary distribution.}
	\label{Rstat}
\end{figure}

\begin{figure}[h]
  \centering
  \subfigure[] {\includegraphics[scale=0.27]{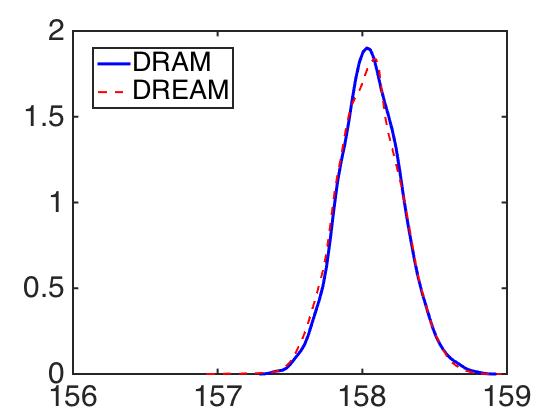}}
  \hspace{-0.1truein}
  \subfigure[]{\includegraphics[scale=0.27]{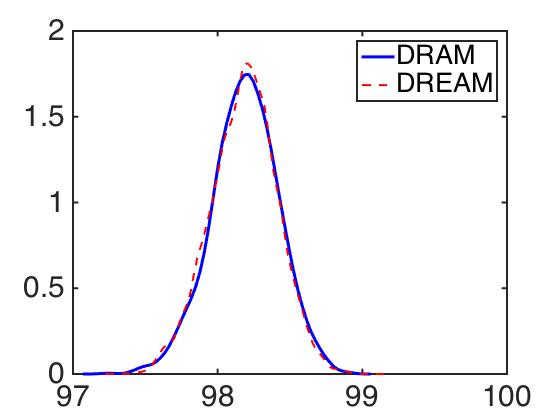}}
  \hspace{-0.1truein}
  \subfigure[]{\raisebox{-4mm}{\includegraphics[scale=0.305]{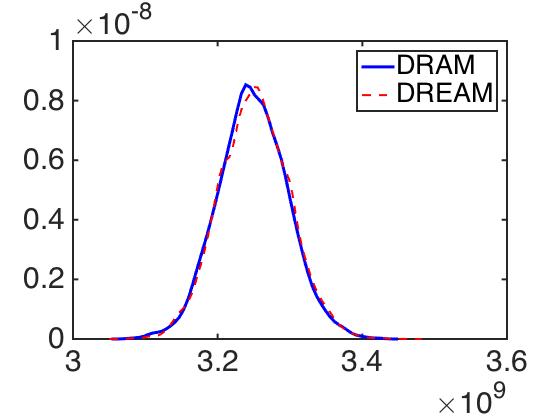}}}
\caption{Comparison of marginal densities for source components (a) $r_s^x$, (b) $r_s^y$, and (c) $S_0)$ obtained with DRAM and DREAM.}
\label{BothDensities}
\end{figure}

Figure \ref{DREAMChains} shows the plot of the ten chains for all three source properties. Truncated DREAM chains displayed in panels (d-f) show good mixing. The stationarity of the chains indicates that they have burned-in and are sampling from the posterior density. This is also confirmed by the plots of the Gelman-Rubin R-statistic in Figure \ref{Rstat}. With the Gelman-Rubin R-statistic values below $1.2$, we conclude that the chains have converged to their stationary distributions. To estimate our source properties, we used the mean value of the final $25\%$ of the chains, which are comprised of samples from the stationary posterior distributions.  The resulting estimates $\hat{{\bf r}}_s^0 = (158.05,98.180)$, and $\hat{S}_0^0 = 3.251 \times 10^9$ compare favorably with the true properties' values.  A comparison of the marginal densities resulting from DRAM and DREAM are given in Figure \ref{BothDensities}. Note that both the radiation source properties estimates and densities produced by DREAM agree with the results from DRAM.

%\begin{figure}[h]
%  \centering
%  \subfigure[] {\includegraphics[scale=0.27]{Figures/DREAM_densityx.jpg}}
%  \hspace{-0.1truein}
%  \subfigure[]{\includegraphics[scale=0.27]{Figures/DREAM_densityy.jpg}}
%  \hspace{-0.1truein}
%  \subfigure[]{\includegraphics[scale=0.27]{Figures/DREAM_densityI0.jpg}}
%\caption{Marginal densities for parameters (a) $x$, (b) $y$, and (c) $I_0$ obtained with DREAM.  }
%\label{DREAMDensities}
%\end{figure}

%\begin{table}[h]
%\begin{center}
%\begin{tabular}[h]{|c|c|}
%\hline
%		& Parameter Estimates  \\
%\hline
%$\hat{x}$	& 158.07	\\
%\hline
%$\hat{y}$	&  98.172 \\
%\hline
%$\hat{I_0}$ &  $3.2510 \times 10^9$ \\
%\hline
%\multicolumn{2}{|c|}{} \\
%\hline
%CPU time (sec) & 37946.18 \\
%\hline
%\hline
%\end{tabular}
%\end{center}
%\caption{Numerical results from DREAM using the Poisson likelihood \eqref{eqn:Poisson_Likelihood}.  The reported parameter estimates are the mean values from the final 25\% of the pooled chain samples.  }
%\label{DREAMresults}
%\end{table} 
%

\section{Conclusion and Future Work} \label{sec:Conclusion}

Searching for radiation material in an urban environment is a challenging problem. By using a simplified piecewise continuous differentiable response model for the complicated urban geometry, we constructed specialized non-smooth likelihood functions based on Poisson distribution. We employed hybrid methodologies by coupling three global optimization methods (1) Simulated Annealing, (2) Particle Swarm, and (3) Genetic Algorithm $+$ Implicit Filtering, the later being a local optimization method designed for objective functions that could be either non-smooth, not everywhere defined, discontinuous, or stochastic.

We investigated several early stopping criteria such as a reduced number of model response runs and a target objective function value to finish the global search. The resulted pseudo-optimal points were then employed to generate narrow sub-domains used by Implicit Filtering to constrain the search. For large numbers of trajectories, all of our proposed parallel global optimization methods were able to approach the proximity of the source components. When coupled with Implicit Filtering, the proposed hybrid techniques successfully identified the source location and intensity. For the design configuration used in this paper, the hybrid techniques decreased by $31 ,~3.1 $ and $3.4 $ times the number of function evaluations required by the Simulated Annealing, Particle Swarm, and Genetic Algorithm methods only. This also translated into smaller computational times and the hybrid algorithms being $17 \times,~2.1 \times $ and $2.5 \times$ faster than the global approaches. Among these coupled methods, the Simulated Annealing + Implicit Filtering was the fastest approach while the Particle Swarm + Implicit Filtering was the most efficient one with the smallest number of model response runs. The estimated radiation source location and intensity were very accurate and similar to those produced by solely global methods.

We also utilized Delayed Rejection Adaptive Metropolis and DiffeRential Evolution Adaptive Metropolis algorithms to verify the results obtained by the hybrid techniques. The Geweke diagnostic and Gelman-Rubin R statistic revealed the convergence of the chains to their target distributions. When compared with the maximum likelihood estimates, the mean of the marginal distributions of the source properties acquired by the Differential Evolution Adaptive Metropolis and the Delayed Rejection Adaptive Metropolis methods are in agreement.
Future work will include testing our algorithms for more complicated scenarios such as those with multiple and potential moving sources. We will also develop computationally efficient methods to determine optimal trajectories of mobile sensors employed for source identification in distributed parameter systems.

%Identification of a moving source for a parabolic equation with constant coefficients was
%attacked from a different perspective in [22]. The authors demonstrate that a knowledge
%of the instantaneous concentration distribution of the dispersion on any bounded open set
%located away from the support of a source is sufficient to estimate a nontrivial subset of the
%actual convex hull of the support of the source which they call the carrier support.
%Estimation of moving point sources in a three-dimensional homogeneous solid in transient
%heat conduction was in investigated in [23]. The identification procedure is based on a
%boundary element method formulation using transient fundamental solutions. The results
%were verified on both simulation and experimental examples. 

\section*{Appendix}
\addcontentsline{toc}{section}{Appendices}
\renewcommand{\thesubsection}{\Alph{subsection}}

Here we will use a more general notation $\pt = [\theta_i]_{i=1,2,3}$  and $\Omega = [l_1,u_1] \times [l_2,u_2] \times [l_3,u_3]$ to represent the source location and intensity and the feasible space. References for all the algorithms are provided in the associated subsections of Sections \ref{sec:hybrid_techniques},\ref{sec:Bayesian_techniques}.

\subsection{ Simulated Annealing} \label{sec:Appendix_SA}

An iteration of the adaptive Simulated Annealing is described in Algorithm \ref{alg::SA_algorithm}. The maximum number of accepted candidates $\max_{\textrm{accept}}$ is set first. Along with it, the initial temperature $\pmb{T}^0$, the reannealing parameter $r_p$, and the first guess $\told = [\theta_i^{\textrm{old}}]_{i=1,2,3}$ of the searched source properties must be defined too. Each  $\theta_i^{\textrm{old}}$ has an associated temperature $T_i$. The components $\theta_i^{(*)},~i=1,2,3$ of the new state point candidate are drawn from uniform proposal distributions $U(\theta_i^{\textrm{old}}-T_i,\theta_i^{\textrm{old}}+T_i),~i=1,2,3$, in case the boundaries constraints are satisfied.

\begin{algorithm}[t!]
\caption{Adaptive Simulated Annealing}\label{alg::SA_algorithm}
\begin{algorithmic}[1]
\State Set $\max_{\textrm{accept}}$ , $\pmb{T}^0$, $r_p$ and $\pmb{k} = [k_i]_{i=1,2,3} ={\bf 1}$. Select initial point $\told \in \Omega$. Set $T_i = T_i^0, i=1,2,3.$
\State Sample $r_i,~i=1,2,3,$ from $U(-1,1)$. Generate a new sample using the proposal function
\begin{equation}\label{eq::SA_proposal_function}
  {\tilde\theta}_i = \toldi + r_i \cdot T_i,~i=1,2,3.
\end{equation}
\If{$\tilde{\pmb{\theta}} = [\tilde \theta_i]_{i=1,2,3}$ is feasible} set
$ \pmb{\theta}^{(*)} = {\tilde{\pmb{\theta}}},$
\Else{ set}
\begin{equation}
  \pmb{\theta}^{(*)} = \alpha\cdot{\bar{\pmb{\theta}}} + (1-\alpha)\told, \textrm{ where }\alpha \sim U(0,1)
  \label{eq::convex_comb_SA}
\end{equation}
 and ${\bar{\pmb{\theta}}} = {\tilde{\pmb{\theta}}} $, except the components violating the constrains which are set to the associated bounds.
\EndIf.
\If
{ $
  J(\pmb{\theta}^{(*)}) < J(\told)
}$
set $ \tnew = \pmb{\theta}^{(*)}$,

\Else { accept $\pmb{\theta}^{(*)}$ with probability} $p_a = \frac{1}{1+\exp{\frac{J(\pmb{\theta}^{(*)})-J(\told)}{\max_i(T_i)}}}.$
\EndIf.
\State Annealing scheme:
$\label{eq::annealing_scheme}
 {T}_i = {T_i}^0 \cdot 0.95^{{k}_i}, i=1,2,3.
$

\If {mod(Number of accepted points so far, $r_p) = 0$} \Comment{Reannealing scheme}
\begin{equation}
k_i = \log\left(\frac{T_i^0}{T_i}\cdot\frac{\max_j(s_j)}{s_i}\right), \textrm{ where }s_i = \left|\frac{J(\tnew + \pmb{e}_i\delta) - J(\tnew)}{\delta}\right|,~i=1,2,3, \label{eqn::SA_objective_perturb}
\end{equation}
\Else { set $\pmb{k} = \pmb{k} + 1$.}
\EndIf.
\State Set $\told = \tnew$ and GO TO step (2).
\end{algorithmic}
\end{algorithm}

For constraint violations, an intermediary sample $\bar{\pmb{\theta}}$ is introduced, replacing the outbound components with the associated bounds. The new proposal is then obtained as a convex combination of $\bar{\pmb{\theta}}$ and $\told$ \eqref{eq::convex_comb_SA} with $\alpha$ being sampled from a uniform distribution $U(0,1)$.

All downhill proposals state points are accepted. In case an uphill candidate is obtained, it is accepted with probability $p_a$. Larger temperatures $T_i,~i=1,2,3,$ increase the chances that uphill candidates are accepted. The algorithm includes an annealing scheme where the temperature values are lowered according to line \ref{eq::annealing_scheme} of the Algorithm \ref{alg::SA_algorithm}. Once the temperatures start to decrease, future points that increase the objective function value are less probable to be accepted. A reannealing scheme is scheduled after each $r_p$ accepted samples. The algorithm stops if the searched source components have remained unchange for a few successive reannealing steps or the predefined maximum number of cost function evaluations has been reached or a preset objective function percentage decrease has been achieved.

\subsection{Particle Swarm} \label{sec:Appendix_PS}

\begin{algorithm}[b!]
\caption{Particle Swarm - Initialization }\label{alg::PS_algorithm_init}
\begin{algorithmic}[1]
\State Select swarm size $P \in \mathbb{N}$ and generate initial state points $[\pmb{\theta}^{\textrm{old}}]^j$ and velocities $[\pmb{v}^{\textrm{old}}]^j,$ $j=1,2,..,P$ such that $[{\theta}^{\textrm{old}}_i]^j,~[{v}^{\textrm{old}}_i]^j \in [l_i,u_i],~i=1,2,3$.
\State Select the minimum neighborhood size $\textrm{minNs}$ and the inertia parameters $W^j \in \mathbb{R},~j=1,2,..,P,~W^j \in [0.1,1.1]$.
\State Set the stall counter $c^j = 0$ for all state points $j=1,2,..,P$.
\State Set the self and social adjustment real variables $y_1$ and $y_2$.
\State Set $N = \textrm{Ns}$.
\end{algorithmic}
\end{algorithm}

\begin{algorithm}[t!]
\caption{Particle Swarm - $j^{\textrm{th}}$ trajectory}\label{alg::PS_algorithm_particle}
\begin{algorithmic}[1]
\State Select $N$ state points other than $j$ to generate the associated neighbourhood. \label{alg_line::loop}
\State Set $\textrm{flag} = \textrm{false}$. Define set $S$ containing all of the $N$ state points. Find the lowest objective function
\begin{equation*}
  {\bf g} = \argmin_{{\pmb{\theta}}^{\textrm{old}} \in S} J({\pmb{\theta}}^{\textrm{old}}) \textrm{ and set } J^{n*} = J({\bf g}).
\end{equation*}
\State Select random vectors ${\bf u}_1$ and ${\bf u}_2$ of size $3$ from the uniform distribution $U(0,1)$. Update the velocity:
\begin{equation}\label{eqn::PS_velocity_formula}
  {\bf v}^{\textrm{new}} = W\cdot {\bf v}^{\textrm{old}} + y_1 \cdot {\bf u}_1.*({\bf p}-{\pmb{\theta}}^{\textrm{old}}) + y_2 \cdot {\bf u}_2.*({\bf g}-{\pmb{\theta}}^{\textrm{old}}).
\end{equation}
%where ${\bf p}$ represents the state point corresponding to the lowest objective function value $J^*$ among all the previous positions of the trajectory.
\State Update the position
\begin{equation}\label{eqn::PS_position_formula}
{\pmb{\theta}}^{\textrm{new}} = {\pmb{\theta}}^{\textrm{old}} + {\bf v}^{\textrm{new}}.
\end{equation}
\State Enforce the bounds. If any component of ${\pmb{\theta}}^{\textrm{new}}$ is outside a bound, set it equal to that bound.
\If{$J({\pmb{\theta}}^{\textrm{new}}) < J^*$} ${\bf p} = {\pmb{\theta}}^{\textrm{new}},~J^*=J({\pmb{\theta}}^{\textrm{new}})$
\EndIf.
\If{$J({\pmb{\theta}}^{\textrm{new}}) < J^*_b$} $\textrm{flag} = \textrm{true},~ J^*_b = J({\pmb{\theta}}^{\textrm{new}}) \textrm{ and } \pmb{b} = {\pmb{\theta}}^{\textrm{new}},$
where $J^*_b$ corresponds to the smallest objective function in the swarm.
\Else $\textrm{ flag} = \textrm{false}$
\EndIf.
\If{\textrm{flag} = \textrm{true}} set $c = \max(0,c-1)$ and $N = \textrm{Ns}$.
\If{$c < 2$} $W = 2\cdot W$
\EndIf.
\If{$c > 5$} $W = W/2 $ and ensure that $W$ is inside the bounds.
\EndIf.
\Else $~$ set $c = c + 1$, $N = \min(N+\textrm{Ns},P)$
\EndIf.
\State Set $\pmb{\theta}^{\textrm{old}} = \pmb{\theta}^{\textrm{new}}$ and $\pmb{v}^{\textrm{old}} = \pmb{v}^{\textrm{new}}$ and GO TO step \ref{alg_line::loop}.
\end{algorithmic}
\end{algorithm}

%The $j^{\textrm{th}}$ state point trajectory evolution depends on the previous best position $\pmb{p}^j$ of the trajectory and the best position $\pmb{g}^j$ among a subset of state points other than $j$.

The initialization stage of Particle Swarm is described in Algorithm \ref{alg::PS_algorithm_init}. The algorithm starts by selecting the population size of the swarm denoted by $P$.  Initially, the state positions $[\pmb{\theta}^{\textrm{old}}]^j$ and velocities $[\pmb{v}^{\textrm{old}}]^j,$ $j=1,2,..,P$ are randomly selected from uniform distributions; i.e., $[{\theta}^{\textrm{old}}_i]^j,~[{v}^{\textrm{old}}_i]^j \sim U[l_i,u_i],~i=1,2,3$. Each state point has an associated neighborhood of size $N=\textrm{Ns}$ influencing its future trajectory. Other parameters of the algorithm must be selected too, such as the inertia parameters $W^j \in \mathbb{R}$ and stall counter $c^j$ for $j=1,2,..,P$. These parameters influence the space search.

The evolution of the space point $j^{\textrm{th}}$ from the current state to the next one is described in Algorithm \ref{alg::PS_algorithm_particle}. The index notation is dropped. The proposal function depends on a two steps formula. First, the velocity  ${\bf v}^{\textrm{new}}$ is adjusted via equation \eqref{eqn::PS_velocity_formula} while, in the second phase, the new state is obtained by adding the newly generated velocity to its previous position \eqref{eqn::PS_position_formula}. The weights $y_1$ and $y_2$ denote the self and social adjustment coefficients steering the search towards either the state point ${\bf p}$ or its neighbours ${\bf g}$ best position. By $.*$ we denote the Hadamard product.

A successful replacement of the best state point position ${\bf b}$ among the entire population ensures a change in the inertia parameter $W$ while a failure leads to a larger neighborhood selection and maintains $W$ constant. Finally the new proposals are set to replace the current ones for the next iteration. The algorithm stops when the relative change in the lowest objective function value $J^*_b$ over a range of predefined number of iterations is smaller than a specified tolerance, the maximum number of iterations is reached, or a preset objective function percentage decrease has been achieved.

\subsection{Genetic Algorithm} \label{sec:Appendix_GA}

\begin{algorithm}[t!]
\caption{Genetic Algorithm}\label{alg::GA_algorithm}
\begin{algorithmic}[1]
\State Select the population size $P$. Choose the initial state points $[\pmb{\theta}^{\textrm{old}}]^j,~j=1,2,..,P$ such that $[{\theta}^{\textrm{old}}_i]^j,~\sim U(l_i,u_i),~i=1,2,3$. Select the elite crossover and mutation fractions $r_e,~r_c,~r_m \in \mathbb{N},~r_e+r_c+r_m = P$.

\State Score each point of the current population by computing its fitness value $J([\pmb{\theta}^{\textrm{old}}]^j),~j=1,2,..,P$. \label{line::GA_alg_iter_start}

\State Select parent points based on their fitness score by uniformly sampling from uniform distribution $U(0,1)$. Higher scores receive a larger portion of the line $[0,1]$.
%Out of the entire current population, select parents points based on their fitness score. This is done by associating a section of the line between $[0,1]$ to each member proportional to its corresponding fitness value. Then samples from the uniform distribution $U(0,1)$ select the states used for the constructing the next generation.

\State Selection of $[\pmb{\theta}^{\textrm{new}}]^j,~j=1,2,..,P$ is done by using elite individuals, mutation and crossover.

\State Elite: $r_e$ state points of the population having the lower fitness values are passed to the next generation
\begin{equation}
\label{eqn::elite_operator}
  [\pmb{\theta}^{\textrm{new}}]^{l} = [\pmb{\theta}^{\textrm{old}}]^{l},~l=l_1,l_2,..,l_{r_e}.
\end{equation}

\State Crossover: $r_c$ state points are created by combining entries of the parents using uniform random weights:
\begin{equation}
\label{eqn::crossover_operator}
  [\pmb{\theta}^{\textrm{new}}]^{s} = \pmb{\lambda}.*[\pmb{\theta}^{\textrm{old}}]^{j_1} + (\pmb{1}-\pmb{\lambda}).* [\pmb{\theta}^{\textrm{old}}]^{j_2},~s=s_1,s_2,..,s_c,
\end{equation}
where components ${\lambda}_i \sim U(0,1),~i=1,2,3,$ while $j_1$ and $j_2$ are uniformly drawn from the parent points.

\State Mutation: $r_m$ points are generated by making random changes to a single parent point:
\begin{equation}
\label{eqn::mutation_operator}
  [\pmb{\theta}^{\textrm{new}}]^t = \pmb{\lambda}.*\pmb{\theta}^{\textrm{old}}_t + (\pmb{1}-\pmb{\lambda}).*\pmb{\varepsilon},~t = t_1,~t_2,..,t_{r_m},
\end{equation}
where $\pmb{\varepsilon}$ and $\pmb{\lambda}$ components are drawn from uniform distribution $U(l_i,u_i),~i=1,2,3$ and $U(0,1)$.

\State Set $[\pmb{\theta}^{\textrm{old}}]^{j} = [\pmb{\theta}^{\textrm{new}}]^{j},~j=1,2,..,P$ and GO TO \ref{line::GA_alg_iter_start}.
\end{algorithmic}
\end{algorithm}

The procedure described in Algorithm \ref{alg::GA_algorithm}, starts by selecting the population size $P$ and the first generation of state points $[\pmb{\theta}^{\textrm{old}}]^j,~j=1,2,..,P$. The algorithm then creates a sequence of new points iteratively referred to as children from the current points known as parents. In each generation, the objective function of every state point in the population is evaluated. The points associated with lower objective function values have higher chances to be considered parents. Moreover, a point can be selected more than once as a parent, in which case it will contribute its genes to more than one child.

The elite operator \eqref{eqn::elite_operator} guarantees that a number of state points with the smallest objective function values in the current generation will be passed to the next generation. In addition, the algorithm creates crossover individuals \eqref{eqn::crossover_operator} by combining pairs of parent points in the current population. The mutation operator \eqref{eqn::mutation_operator} also generates new points by randomly changing the components of some of the current parents.

The algorithm stops when either the number of generations or function evaluations is larger than some prescribed values or the current lowest objective function is smaller than some predefined threshold.

\subsection{Implicit Filtering}  \label{sec:Appendix_IF}

Implicit Filtering may be described simply as a series of outer and inner iterations. The outer iteration, described in Algorithm \ref{alg::Implicit_Filtering_outer}, simply verifies if the stencil size $h$ and number of function evaluations are maintained under some prescribed values. The number of stencil directions is taken as twice the number of $\pt$ components. The parameter $\beta$ defines the backtracking line search step size of the inner iteration. Parameters $\textrm{maxit}$ and $\textrm{maxitarm}$ limit the number of loops in the inner iterations and number of step size reductions within the line search, while $\tau$ controls the modified projected Quasi-Newton stopping criterion. The optimization will terminate when the updated stencil size $h$ is smaller then some predefined threshold $h_{\min}$ or if the budget of function evaluations is exceeded.

\begin{algorithm}[t!]
\caption{Implicit Filtering - outer iteration}\label{alg::Implicit_Filtering_outer}
\begin{algorithmic}[1]
\State Select initial state $\pmb{\theta}$, stencil size $h$, $h_{min}$, stencil directions $V$, $\textrm{budget}$, $\textrm{maxit}$, $\textrm{maxitarm}$, $\tau$, $\beta$.
\State Set $J_{base} = J(\pmb{\theta})$ and $J_{count} = 1$.
 \While{$J_{count}\leq \textrm{budget} \textrm{ and } h \geq h_{min}$}
 \State inner loop $\to$ $J_{base},~\pmb{\theta},~icount$
 \State Set $J_{count} = J_{count} + icount$ and $h = h/2$.
 \EndWhile
\end{algorithmic}
\end{algorithm}

\begin{algorithm}[t!]
\caption{Implicit Filtering - inner iteration}\label{alg::Implicit_Filtering_inner}
\begin{algorithmic}[1]
%\State
\State Set $p=1$ and $\varepsilon = 10^{-6}$. Compute $J_{base} = J(\pmb{\theta})$ and evaluate stencil gradient $\nabla J(\pmb{\theta},V,h)$.
 \While{$p\leq \textrm{maxit} \textrm{ and } \| \pmb{\theta} - \mathcal{P}\left(\pmb{\theta}-\nabla J(\pmb{\theta},V,h)\right) \| \geq \tau \cdot h$}  \label{eqn::proj_QN_stop_crit}
 \State $j=1$. \label{eqn::probe_stencil1}
 \For{i=1,2,..,K} \label{eqn::probe_stencil2}
 \If{$l_i\leq{\theta}_i \leq u_i,\textrm{ for all }  i=1,2,3$}  $J_j = J(\pmb{\theta} + h{\pmb{v}_i})$ and $j=j+1$. \label{eqn::probe_stencil4}
 \EndIf. \label{eqn::probe_stencil5}
 \EndFor. \label{eqn::probe_stencil6}
 \State Find $i^{*}$ such that $J(\pmb{\theta} + h\pmb{v}_i^{*}) = \min_{i} J_i.$
 \State Set $\pt_{\min} = \pt + h\pmb{v}_i^{*}$ and $J_{\min} = J(\pt_{\min})$.
 \If{$J_{\min} > J_{base}$} Terminate inner loop on stencil failure.
 \EndIf.
 \State Update the model Hessian $\mathcal{R}$ and solve
 \begin{equation}\label{eqn::QN_direction}
   \mathcal{R}d = - \nabla J(\pt,V,h).
 \end{equation}
 \State Set $\lambda = \beta$ and $\pt^{\mathcal{R},\varepsilon}(\lambda) = \mathcal{P}(\pt + \lambda d)$.
 \State Backtracking line search: Find the smallest integer $m \leq \textrm{maxitarm}$ such that
 \begin{equation}\label{eqn::IF_line_search}
   J(\pt^{\mathcal{R},\varepsilon}(\lambda)) < J(\pt),~\textrm{where } \lambda=\beta^m.
 \end{equation}
 \If{Line search succeeds} set $\pt = \pt^{\mathcal{R},\varepsilon}(\lambda))$
 \Else  $~$set $\pt = \pt_{\min}$
 \EndIf.
 \State Set $J_{base} = J(\pt)$ and evaluate $\nabla J(\pt,V,h)$. Set $p = p + 1$ and update $icount$.
 \EndWhile
\end{algorithmic}
\end{algorithm}

The inner iteration, described in Algorithm  \ref{alg::Implicit_Filtering_inner}, starts by setting the loop counter $p$ and parameter $\varepsilon$ associated with a relaxed $\varepsilon-$ binding set \cite{kelley2011implicit}. Next, the finite difference stencil gradient $\nabla f(\pt,V,h)$ and the stopping criterion of the projected Quasi-Newton method are evaluated. Here, $\mathcal{P}$ denotes the projection onto the feasible space
\begin{equation}\label{eqn::projection_definition}
     \mathcal{P}  : \mathbb{R} \times \mathbb{R} \times \mathbb{R} \to \Omega,~~
     \mathcal{P}(\pmb{x})   = \left[\max\left(l_i, \min(x_i,u_i)\right)_{i=1,2,3}\right]^T.
\end{equation}

Next the algorithm probes the stencil and evaluates the objective function for all the feasible points inside the stencil as seen in lines \ref{eqn::probe_stencil1}-\ref{eqn::probe_stencil6} of Algorithm \ref{alg::Implicit_Filtering_inner}. In case the objective function value at the current point is smaller than all the values anywhere on the stencil, the inner iteration is terminated and it is said that a stencil failure has been encountered.

In case a better point $\pt_{min}$ in the stencil is identified, then the Quasi-Newton method proceeds by approximating the model Hessian at the current point $\pt$ using the projected Broyden-–Fletcher–-Goldfarb–-Shanno updating formula \cite{kelley2011implicit}. Once the new direction $d$ is obtained \eqref{eqn::QN_direction}, a modified backtracking line search procedure starts seeking a step length $\lambda$ that satisfies \eqref{eqn::IF_line_search} in no more than $\textrm{maxitarm}$ steps. The number of the step size reductions is limited and the line search formula modified to accept simple decrease to accommodate non-smooth objective functions. If the line search obtains a downhill point in less than $maxitarm$ steps, we say that the line search succeeded and update the current state $\pt$. Otherwise the actual state is set to $\pt_{\min}$, the stencil point with the smallest objective function. Finally the finite difference approximation of the gradient of the new proposal is computed and the algorithm prepares for a new inner loop by incrementing the counter $p$ and updating the number of function evaluations so far in the inner iteration.

\subsection{Delayed Rejection Adaptive Metropolis} \label{sec:Appendix_DRAM}

\begin{algorithm}[b!]
\caption{Delayed Rejection Adaptive Metropolis   \label{DRAM1}}
\begin{algorithmic}[1]

%\item[(1)] Set design parameters $n_s$, $s_p$, $\sigma_s^2$, and $k_0$ and the number of chain iterates $M$.
\State Set design parameters $s_p$ and $k_0$ and the number of chain iterates $M$.
\State Determine $\pt^0=\argmin_{\pt}\sum_{i=1}^{10}\sum_{j=1}^{10}[v_{ij}-f(D_i,\pt)]^2.$
\State Select positive definite covariance matrix $V_0$ and compute Cholesky decomposition $V_0 = R_0 R_0^T$.
\For{$k=1,...,M$}
\State Sample $\pmb{z}^k \sim \mathcal{N}(0,I)$, where $\pmb{z}^k \in \mathbb{R}^3$ and $I$ is the corresponding identity matrix.
\State Construct candidate
\begin{equation}\label{eqn::DRAM_proposal_function}
  \pt^* = \pt^{k-1}+R_{k-1}^T \pmb{z}^k.
\end{equation}
\State Sample $u_{\alpha} \sim \mathcal{U}(0,1)$. Compute $\alpha(\pt^*|\pt^{k-1}) = \min \left(1, \dfrac{\pi(\pmb{V}|\pt^*)\pi_0(\pt^*)}{\pi(\pmb{\pmb{V}}|\pt^{k-1})\pi_0(\pt^{k-1})}\right)$ using likelihood functions $\pi$ and prior $\pi_0$. \label{eq::DRAM_acceptance_rate1}
\If{$u_{\alpha}<\alpha$} set $\pt^k = \pt^*$ %, $SS_{\theta^k} = SS_{\theta^*}$
\Else $~$ Enter Delayed Rejection Algorithm \ref{DRAM2}.
\EndIf.
\If{$\mod(k,k_0) = 1$} \label{eqn::DRAM_adaptive1} update $V_k = s_p\text{cov}(\pt^0,\pt^1,...,\pt^k)$ and compute decomposition $V_k = R_kR_k^T$
\Else $~V_k = V_{k-1}$. %and $R_k = R_{k-1}$
\EndIf. \label{eqn::DRAM_adaptive2}
\EndFor.
\end{algorithmic}
\end{algorithm}

The basic algorithm for DRAM is given in Algorithm \ref{DRAM1} with the delayed rejection component described in Algorithm \ref{DRAM2}. Initially the DRAM algorithm requires the selection of a covariance design parameter $s_p$, the adaptation interval length $k_0$ and the maximum number of allowed chain iterates $M$. The parameter $k_0$ determines when the covariance matrix $V_k$ of the chain should be updated. The choice of $k_0$ is critical for a good balance mixing in the initial stages and for generating non-singular covariance matrices. For our numerical experiments, the diagonal elements of the initial covariance matrix were selected $[V_0]_{ii} = |\theta_i^0\cdot 0.05|^2,~i=1,2,3$ and the length of the adaptation interval was chosen $k_0 = 100$. To scale the updated covariance matrix we set $s_p = \frac{2.38^2}{3}$.

\begin{algorithm}[t!]
\caption{Delayed Rejection Component of DRAM \cite{Haario,smith2014uncertainty}} \label{DRAM2}
\begin{algorithmic}[1]
\State Set the design parameter $\gamma_2 < 1$.  We select $\gamma_2 = \frac{1}{5}$.
\State Sample $\pmb{z}^k\sim \mathcal{N}(0,I)$ and construct second-stage candidate
\begin{equation*}
  \pt^{*2} = \pt^{k-1}+\gamma_2R^T_{k-1}\pmb{z}^k.
\end{equation*}
	
\State Sample $u_{\alpha_2}\sim \mathcal{U}(0,1)$ and compute

 \begin{equation}\label{eq::DRAM_acceptance_rate2}
   \alpha_2(\pt^{*2}|\pt^{k-1},\pt^*) = \min\left(1,\frac{\pi(\pmb{V}|\pt^{*2})\pi_0(\pt^{*2})\hat{J}(\pt^*|\pt^{*2})[1-\alpha(\pt^*|\pt^{*2})]}{\pi(\pmb{V}|\pt^{k-1})\pi_0(\pt^{k-1})
   \hat{J}(\pt^*|\pt^{k-1})[1-\alpha(\pt^*|\pt^{k-1})]}\right).
 \end{equation}
Here $\hat{J}$ is the proposal, or jumping, distribution used in Algorithm \ref{DRAM1}; i.e.,
\begin{equation*}
  \hat{J}(\pt^a|\pt^b) = \frac{1}{\sqrt{(2\pi)^3|V|}}\exp(-\frac{1}{2}[(\pt^a-\pt^b)V^{-1}(\pt^a-\pt^b)^T],
\end{equation*}
where $\pt^a,~\pt^b$ are general samples with the corresponding covariance matrix $V$ and $|V|$ denotes its gradient.
\If{$u_{\alpha_2}<\alpha_2$} set $\pt^k=\pt^{*2}$  %, $SS_{\theta^k} = SS_{\theta^{*2}}$
\Else $~$set $\pt^k = \pt^{k-1}$ % , $SS_{\theta^k} = SS_{\theta^{k-1}}$
\EndIf.
\end{algorithmic}
\end{algorithm}

The $k^{\textrm{th}}$ state point candidate is obtained by sampling the normal distribution with mean $\pt^{k-1}$ and covariance $V_{k-1}$. Initially, the proposal is accepted with probability $\alpha(\pt^*|\pt^{k-1})$ defined in line \ref{eq::DRAM_acceptance_rate1} of the Algorithm \ref{DRAM1}. In case of rejection, the delayed rejection component will generate an alternative candidate $\pt^{*2}$ instead of retaining the previous state point $\pt^{k-1}$. The second proposal $\pt^{*2}$ is chosen from a narrower distribution $N(\pt^{k-1},\gamma_2 V_{k-1})$ since $\gamma_2<1$ to improve mixing.

By choosing the probability of accepting the second candidate as in equation \eqref{eq::DRAM_acceptance_rate2}, it is guaranteed that the detailed balance condition is satisfied. In consequence, by sampling long enough, we should be able to generate state points from the stationary distribution as long as the diminishing adaptation and bounded convergence conditions are satisfied \cite{andrieu2008tutorial,haario2001adaptive,roberts2009examples}.

\subsection{Differential Evolution Adaptive Metropolis} \label{sec:Appendix_DREAM}

The DREAM method can be interpreted as a collection of $P$ chains $\pt^j,~j=1,2,..,P$ simultaneously ran in parallel with samples for each chain extracted using the proposal function defined in \eqref{eqn::proposal_function_DREAM}. Using more than two members for new candidates increases diversity. For our simulations, we used $\delta = 3$ and the value of the jump-size was selected to be $\gamma(\delta) = \frac{2.38}{\sqrt(6)}$.

The algorithm takes advantage of a randomized subspace sampling strategy and accepts as proposals only those components $\theta^{*j}_i$ satisfying the probability scheme described in \eqref{eqn::subspace_sampling_strategy_DREAM}. This is especially useful for high-dimensional feasible spaces. The distribution of crossover probabilities $CR$ \cite{Vrugt} is computed during the burn-in period and favors large jumps over smaller ones in each of the $P$ chains thus exploring the search space faster.

\begin{algorithm}[t!]
\caption{DREAM}\label{alg::DREAM}
\begin{algorithmic}[1]
\State Select the number of chains $P$ and draw an initial population $\{\pt^j,~j=1,2,..,P\}$ using a uniform prior distribution, i.e. $[{\theta}_i]^j,~\sim U(l_i,r_i),~i=1,2,3.$ Set the number of pairs $\delta$ and $\gamma(\delta)$.
\For{j=1:P}  \label{alg::line_loop_DREAM}
\State Generate a new candidate using the proposal function
\begin{equation}\label{eqn::proposal_function_DREAM}
  \pt^{*j} = \pt^j+\left(\mathcal{I} + E\right)\gamma(\delta,d)\left[\sum_{i=1}^{\delta}\pt^{j_1(i)}-\sum_{n=1}^{\delta}\pt^{j_2(n)}\right] + \pmb{\varepsilon},
\end{equation}
where $j_1(i),~j_2(n) \in \{1,2,..,P\}$ with $j_1(i)\neq j_2(n) \neq j,$ for $i,n=1,2,..,\delta$. The components of matrix $E$ and vector $\pmb{\varepsilon}$ are realization of uniform and normal distributions, i.e. $e_{l,k} \sim U(-b,b),~l,k=1,2,3$ and ${\varepsilon}_l \in \mathcal{N}(0,b^*),~l=1,2,3,$, where $b$ and $b^*$ are smaller than the variance of the posterior density. The size of the identity matrix $\mathcal{I}$ is $3$ corresponding to the number of the radiation source properties.

\State A randomized subspace sampling scheme does not allow sampling all components simultaneously
\begin{equation}\label{eqn::subspace_sampling_strategy_DREAM}
  \theta_i^{**j} = \left \{
  \begin{tabular}{c}
  $\theta_i^j,~\textrm{if } u \leq 1 - CR,~\textrm{Set },$ \\
  $\theta_i^{*j},~\textrm{otherwise                           }$,
  \end{tabular}
\right.
\end{equation}
where $CR$ denotes a crossover probability and $u$ is drawn from uniform distribution $\mathcal{U}(0,1)$.

\State Compute $\pt^{**j}$ and Metropolis acceptance probability:
$
  \alpha = \min\left[1,\frac{\pi(\pmb{V}|\pt^{**j})\cdot\pi_0(\pt^{**j})}{\pi(\pmb{V}|\pt^j)\cdot\pi_0(\pt^j)}\right].
$
\State Sample $u \in \mathcal{U}(0,1)$
\If{$\alpha > u$} \label{alg::line_AM_DREAM} set $\pt^j = \pt^{**j}$
\Else $~$ Enter Delayed Rejection Algorithm .
\EndIf.
\EndFor
\State GOTO line \ref{alg::line_loop_DREAM}.
\end{algorithmic}
\end{algorithm}

The candidate $\pt^{**j}$ is accepted with the Metropolis acceptance probability rate computed using the likelihood function \eqref{eqn:Poisson_Likelihood} and a prior distribution $\pi_0$. In case the candidate is rejected, a delayed rejection stage similar to the one proposed by DRAM is utilized.The second candidate $\pt^{***j}$ is accepted with a probability rate similar with the one defined in \eqref{eq::DRAM_acceptance_rate2} such that the detailed balance of the $j^{th}$ chain is preserved.

To increase the performances of the sampler, the outliers produced by the chains are replaced with the current best state of all chains using the standard Inter-Quartile-Range statistic. This is done during the burn-in phase since this step does not preserve the detailed balance. The stopping criterion of the scheme relies on the Gelman-Rubin convergence diagnostic \cite{gelman1992inference} computed using the last $50\%$ percent of the samples in each chain.

%-------------------------------------------------------------------------
\section*{Acknowledgements}
%-------------------------------------------------------------------------
This research was supported by the Department of Energy National Nuclear Security Administration NNSA Consortium for Nonproliferation Enabling Capabilities (CNEC) under the Award Number DE-NA0002576. R\u{a}zvan \c{S}tef\u{a}nescu thanks Prof. Michael Navon for his valuable suggestions on the current research topic, and Mallory McMahon for her contribution to improve the body text of the manuscript.

\newpage

%\section*{\refname}
\bibliographystyle{plainnat}
\bibliography{references,jason}
%\begin{thebibliography}{ }
%
%%1
%\bibitem{Haario} H. Haario, M. Laine, A. Mira, and E. Saksman, ``DRAM: Efficient adaptive MCMC,'' \textit{Statistics and Computing}, 16(4), pp. 339-354, 2006.
%
%%2
%\bibitem{Smith} R.C. Smith, {\it Uncertainty Quantification: Theory, Implementation, and Applications}, SIAM, Philadelphia, PA, 2014.
%
%%3
%\bibitem{Vrugt}  J.A. Vrugt., C.J.F. ter Braak, C.G.H. Diks, B.A. Robinson, J.M. Hyman, D. Higdon, ``Accelerating Markov chain Monte Carlo simulation by differential evolution with self-adaptive randomized subspace sampling," {\it International Journal of Nonlinear Sciences and Numerical Simulation} 10(3), pp. 273-290. doi: 10.1515/IJNSNS.2009.10.3.273, 2009
%
%
%
%\end{thebibliography}
\end{document}